\documentclass[twocolumn]{aastex631}

\usepackage{bm}
\usepackage{amsmath}
\usepackage{comment}
\usepackage{outlines}

\newcommand{\de}{\text{d}}
\newcommand{\Msun}{\text{M}_{\odot}}
\newcommand{\kmsec}{\text{km}\,\text{s}^{-1}}

\newcommand{\kpc}{\text{kpc}}
\newcommand{\pc}{\text{pc}} 

\newcommand{\Myr}{\text{Myr}}

\newcommand{\Msunppcc}{\Msun\pc^{-3}}

\newcommand{\vlos}{v_\mathrm{l.o.s.}}

\newcommand{\popp}{\boldsymbol{\Psi}}

\definecolor{jhcolor}{rgb}{0.1,0.5,0.60}

\begin{document}

\title{The phase spiral's origin and evolution: indications from its varying properties across the Milky Way disk}

\author[0000-0001-5686-3743]{Axel Widmark}
\affiliation{Columbia University, 116th and Broadway, \\ New York, NY 10027 USA}
\affiliation{Stockholm University and The Oskar Klein Centre for Cosmoparticle Physics, \\
Alba Nova, 10691 Stockholm, Sweden}

\author[0000-0001-6584-6144]{Kiyan Tavangar}
\affiliation{Columbia University, 116th and Broadway, \\ New York, NY 10027 USA}

\author{Josh Kalish}
\affiliation{Columbia University, 116th and Broadway, \\ New York, NY 10027 USA}

\author[0000-0001-6244-6727]{Kathryn V. Johnston}
\affiliation{Columbia University, 116th and Broadway, \\ New York, NY 10027 USA}

\author[0000-0001-8917-1532]{Jason A. S. Hunt}
\affiliation{School of Mathematics \& Physics, University of Surrey, \\Stag Hill, Guildford, GU2 7XH, UK}

\correspondingauthor{Axel Widmark} 
\email{axel.widmark@fysik.su.se}

\begin{abstract}
The phase spiral is a perturbation to the vertical phase-space distribution of stars in the Milky Way disk. We study the phase spiral's properties and how they vary with spatial position, in order to constrain its origin and evolution, as well as properties of the disk itself. We produce high resolution maps using two complementary data processing schemes: (a) we bin the \emph{Gaia} proper motion sample in a disk parallel spatial grid, reaching distances up to 4~kpc; (b) we bin the spatially nearby line-of-sight velocity sample in terms of disk parallel orbital parameters. We find complex structure, most significantly with respect to Galactocentric radius and guiding radius, but also in Galactic azimuth and epicyclic action and phase.
We find that spiral winding and rotation phase vary smoothly across the disk, with close-to-flat radial profiles. This uniform structure, in particular for the rotation phase, indicates that the phase spiral was sourced by one or many global perturbations. Curiously, this also implies that the winding time has a strong slope with respect to Galactocentric radius, with low values for the inner disk.
\end{abstract}



\section{Introduction}
\label{sec:intro}

The astrometric \textit{Gaia} mission \citep{Gaia2016mission} has revealed a Milky Way in disequilibrium. One striking time-varying dynamical feature is the ``phase spiral'' discovered by \cite{Antoja2018}. It is a spiral pattern in the phase-space of vertical position and velocity of stars in the Galactic disk. It can be observed in stellar number density, average velocity in directions parallel to the disk-plane, and stellar metallicity abundances \citep{Frankel2024}. This spiral is evidence for a past perturbation to the Galactic disk, which is now phase mixing back towards equilibrium.

The origin of the phase spiral is not known, but several formation mechanisms have been proposed (see \citealt{Hunt2025} for a thorough review). The most widely considered model, first proposed in the \cite{Antoja2018} discovery article, is that the spiral was sourced by a direct perturbation of the disk by the Sagittarius dwarf galaxy \citep[see also, e.g.,][]{Laporte2019,Hunt2021}. Other mechanisms include disk-internal perturbations from the Galactic bar or transient spiral structure \citep[e.g.][]{Khoperskov2019,Hunt2022,Li2023}, or a continual sourcing of spirals due to a dark matter wake \citep{Grand2023}. \cite{Tremaine2023} presented a tantalizing possibility that the phase spiral could be sourced from stochastic noise in the gravitational potential, for example due to dark matter halo substructure (see also \citealt{Gilman2025}). It may be that the spiral is not sourced by a single dominant mechanism, but perhaps some combination of the hypotheses listed above.

The phase spiral's morphology, for example its amplitude, winding, and rotation phase, can help us differentiate between different formation scenarios, as well as constrain properties of the Galactic disk itself. For example, the correlation lengths of spiral properties could distinguish local (i.e. spatially small-scale) and global perturbation scenarios, where the latter would give rise to more uniform structure.

In this effort, many recent articles have mapped the properties of the phase spiral across the disk. Observing the spiral in physical space is challenging due to selection effects from dust and stellar crowding, which are especially severe close to the disk mid-plane. Most previous studies circumvent such difficulties and instead use the well-observed spatially local volume decomposed in orbital parameters, typically parametrized with actions and their conjugate angles. In such analyses, we stress that it is important to be cognizant of the selection effects that are inherent to the local volume. For example, nearby stars with a high angular momentum must, by construction, also have high radial action and currently be close to their orbital pericenter; as such, they are not necessarily representative of high angular momentum stars in general.

It was recognized early on that the phase spiral varies significantly with Galactocentric radius ($R$; see e.g. \citealt{Laporte2019,Xu2020,GaiaCollab2023,Antoja2023}), for example in terms of its axis ratio (width in height relative to vertical velocity), due to the disk surface density decreasing with $R$. 
In fact, the shape of the spiral can be used to directly and precisely infer the disk surface density \citep{Widmark-spiral-I,Widmark-spiral-II,Widmark-spiral-IV,Widmark-spiral-III}.
The spiral has also been shown to vary with Galactic azimuth or azimuthal phase angle ($\phi$ or $\theta_\phi$, which are equivalent for circular orbits), most significantly in terms of the spiral's rotation phase (i.e. orientation in the vertical phase-space plane; see e.g. \citealt{Antoja2018,Hunt2022,Darragh-Ford2023,Alinder2023,Alinder2024}).
The spatially local sample is most commonly decomposed in terms of guiding radius ($R_g$, or angular momentum, $L_z$, equivalently; see, e.g., \citealt{Bland-Hawthorn2019,Khanna2019,Li2021,Gandhi2022,Frankel2023,Antoja2023}). A focal point of many studies is the phase spiral's winding (or pitch angle, equivalently), which is directly related to its perturbation time under some simplifying assumptions, inferred to be in the range of roughly 0.2--1~Gyr for a guiding radius in the range of 6--12~kpc. These inferred times are not perfectly consistent between different orbital parameters, but have a wave-like pattern in guiding radius \citep{Antoja2023}, with additional structure in azimuthal phase \citep[$\theta_\phi$, ][]{Darragh-Ford2023}, and in epicylic action \citep[$J_R$, ][]{Frankel2023}. The precise values for the inferred times vary between studies, likely owing to differences in modeling and the assumed Galactic potential and solar motion, but they do agree qualitatively in terms of the general features.

\cite{Hunt2022} discovered two-armed phase spirals at low $R_g$, which are linked to a symmetric breathing mode \citep[see also][]{Banik2023,Widrow2023} rather than an anti-symmetric bending mode associated with one-armed spirals \citep[e.g.][]{Darling2019}. They further showed that this can occur when an internally excited breathing mode (from a bar or spiral arms) and an externally induced bending mode (from a satellite interaction) overlap. Alternatively, an inner Galaxy breathing mode could arise from transient spiral arms induced by the perturbing satellite \citep{Asano2025}, or by two overlapping one-armed spirals \citep{Li2025}.

The phase spiral is often studied using the spatially local sample, with a binning in one or two phase-space dimensions (see references above). However, a disk plane parallel orbit is only fully described with four parameters, and a complete picture necessitates going beyond the spatially local disk. While distant disk regions are difficult to study due to selection effects, \cite{Widmark-spiral-III} demonstrated that the \emph{Gaia} proper motion sample, which is significantly deeper and less affected by stellar crowding than the line-of-sight velocity measurements, can resolve phase spirals at distances of several kilo-parsecs.

In this work, in an effort to produce a more complete view of the phase spiral's properties, we use two binning schemes: with the proper motion sample, we bin the data in terms of spatial position; with the line-of-sight velocity sample, we bin the data in terms of their disk plane parallel orbits. These two schemes are complimentary; by comparing the two we can draw more robust conclusions about the origin and evolution of the phase spiral.

\section{Coordinate system}
\label{sec:coords}

We use a Cartesian coordinate system, centered on the Sun's location and rest frame, written $\boldsymbol{X} \equiv \{X,Y,Z\}$. The three coordinate axes are pointing in the directions of the Galactic center, of Galactic rotation, and towards the Galactic north, respectively. The velocities along the same axes are written $\boldsymbol{V} \equiv \{U,V,W\}$. The angles of Galactic longitude ($l$) and latitude ($b$) are related to the Cartesian coordinates through
\begin{equation}\label{eq:l_and_b}
\begin{split}
    & \cos(l) = X / \sqrt{X^2+Y^2}, \\
    & \sin(l) = Y / \sqrt{X^2+Y^2}, \\
    & \sin(b) = Z / \sqrt{X^2+Y^2+Z^2}.
\end{split}
\end{equation}

The height with respect to the Galactic plane, also referred to as vertical position, is written as
\begin{equation}\label{eq:Z_to_z}
    z = Z + Z_\odot,
\end{equation}
where $Z_\odot$ is the height of the Sun relative to the stellar disk mid-plane. In this work, we assume a flat disk plane and $Z_\odot = 20~\pc$ (roughly consistent with, e.g., \citealt{Juric2005,Yao2017,Bennett2019}). 

Another crucial quantity in this work is the vertical velocity in the disk's rest frame, written
\begin{equation}\label{eq:W_to_w}
    w = W + W_\odot.
\end{equation}
For the Sun's velocity in the disk rest frame, we use $\boldsymbol{V}_\odot = \{11.1,\, 12.24,\, 7.25\}~\kmsec$ \citep{Schonrich2010}.

We use a Galactocentric radius ($R$) equal to
\begin{equation}
    R = \sqrt{(R_\odot - X)^2 + Y^2},
\end{equation}
where $R_\odot = 8178~\pc$ \citep{GRAVITY2019} is the approximate distance between the Sun and the Galactic center. The Galactic azimuth, written $\phi$, is defined by the relations
\begin{equation}
    \begin{split}
        \cos(\phi) &= \frac{R_\odot - X}{R}, \\
        \sin(\phi) &= \frac{Y}{R},
    \end{split}
\end{equation}
which is zero-valued for the solar position. The disk-plane parallel velocity components, meaning radial and azimuthal in the Galactic rest frame, are given by
\begin{equation}
\begin{split}
    v_R & = (V + V_\odot + v_{c,\odot}) \sin(\phi) - (U+U_\odot) \cos(\phi), \\
    v_\phi & = (V + V_\odot + v_{c,\odot}) \cos(\phi) + (U+U_\odot) \sin(\phi),
\end{split}
\end{equation}
where $v_{c,\odot} = 234~\kmsec$ \citep[consistent with, e.g.,][]{Zhou2023,Ou2024}.

The analysis in this work is not very sensitive to the chosen values for the Sun's position and Milky Way rotational velocity curve. Other reasonable parameter choices yield the same general results and conclusions.

\section{Data and sample construction}
\label{sec:data}

We use data from the \emph{Gaia} mission's third data release (DR3), supplemented with spectro-astrometric parallax estimates from the XP catalog \citep{Zhang2023} and line-of-sight velocity predictions using Bayesian Neural Networks \citep[BNNs;][]{Naik_RVs_EDR3,Naik_RVs_DR3}. We replace the \emph{Gaia} parallax and parallax uncertainty with XP values for stars that fulfill both of these conditions: (i) the XP parallax uncertainty is smaller than the \emph{Gaia} parallax uncertainty; (ii) the XP ``basic reliability cut'' is met (i.e. $\texttt{quality\_flags} < 8$; see \citealt{Zhang2023} for details).

Where $\vlos$ measurements are missing, we use the mean value of the BNN prediction posterior distributions from \cite{Naik_RVs_DR3}. These predictions are informed by the 3d spatial position and 2d proper motion of stars, convolved with measurement uncertainties. The predictions have been thoroughly tested, for example with blind predictions based on \emph{Gaia}'s early third data release \citep{Naik_RVs_EDR3}. The velocity predictions were found to be well behaved for stars within a distance of approximately 7~kpc, and for distance precisions below roughly 20~\%; for the application in this work, we are well within these limits.

The uncertainties of $\vlos$ BNN predictions are typically 25--30~$\kmsec$, roughly equal to the stellar velocity dispersion along the line-of-sight, which is rather large. However, the vertical velocity ($w$) is the component of interest when observing the phase spiral, for which the contribution from $\vlos$ is proportional to $\sin b$. Hence, the uncertainty that propagates from $\vlos$ to $w$ is significant only for stars with high $|b|$ (i.e. nearby and at large $|Z|$). For distant data samples, the propagated errors will have a smearing effect at high $|Z|$, reducing the amplitude of the spiral's relative density perturbation, but will not significantly bias the inferred spiral shape, as argued for and tested in \cite{Widmark-spiral-IV,Widmark-spiral-III}.

\subsection{Quality cuts}
\label{sec:quality_cuts}

We make quality cuts to ensure data precision, by only including stars that fulfill these conditions:
\begin{itemize}
    \item a parallax precision larger than 10 ($\varpi / \sigma_\varpi > 10$);
    \item a renormalized unit weight error (\texttt{RUWE}, a quality measure of the \emph{Gaia} catalog) smaller than 1.4;
    \item open cluster membership probability lower than 50~\%, according to the cluster catalog from \cite{Hunt2023}.
\end{itemize}

These quality cuts, in particular the one in parallax precision, cause significant selection effects that are spatially dependent. There is a general and straight-forward trend that the relative data precision and number of statistics deteriorate with distance. This trend sets an upper limit to how far away we can observe the spiral, but is not expected to give rise to significant systematic biases. A more pressing issue is the presence of significant selection effects on smaller spatial scales, particularly prominent within the stellar disk, associated with dust extinction, stellar crowding, and open clusters.

In order to limit selection effects to having a spatial dependence, we do not apply any quality cuts in $w$. However, quality cuts on proper motion are still present implicitly, since parallax precision correlates strongly with proper motion precision. As a result, when calculating a star's vertical velocity, the uncertainty propagated from the parallax is dominant with respect to that of the proper motion.

\subsection{Spatial binning}
\label{sec:spatial_binning}

We divide the Galactic plane into a hexagonal 2d grid, with a spacing of 400~pc, out to a distance of 4~kpc, giving a total of 365 bins. Throughout this article, we use the term ``data sample'' to refer to the stellar sample that falls within the confines of a single bin. The spatially binned data samples are split into two classes:
\begin{itemize}
    \item For nearby data samples, whose mid-point distance is equal to or smaller than 1.6~kpc, we require an available \emph{Gaia} $\vlos$ measurement for all stars.
    \item For distant data samples, beyond 1.6~kpc, we also include the proper motion sample. We use the \emph{Gaia} $\vlos$ measurement if available, and otherwise the BNN $\vlos$ prediction (see the beginning of Section~\ref{sec:data}).
\end{itemize}

\begin{figure*}
    \centering
    \includegraphics[width=0.9\textwidth]{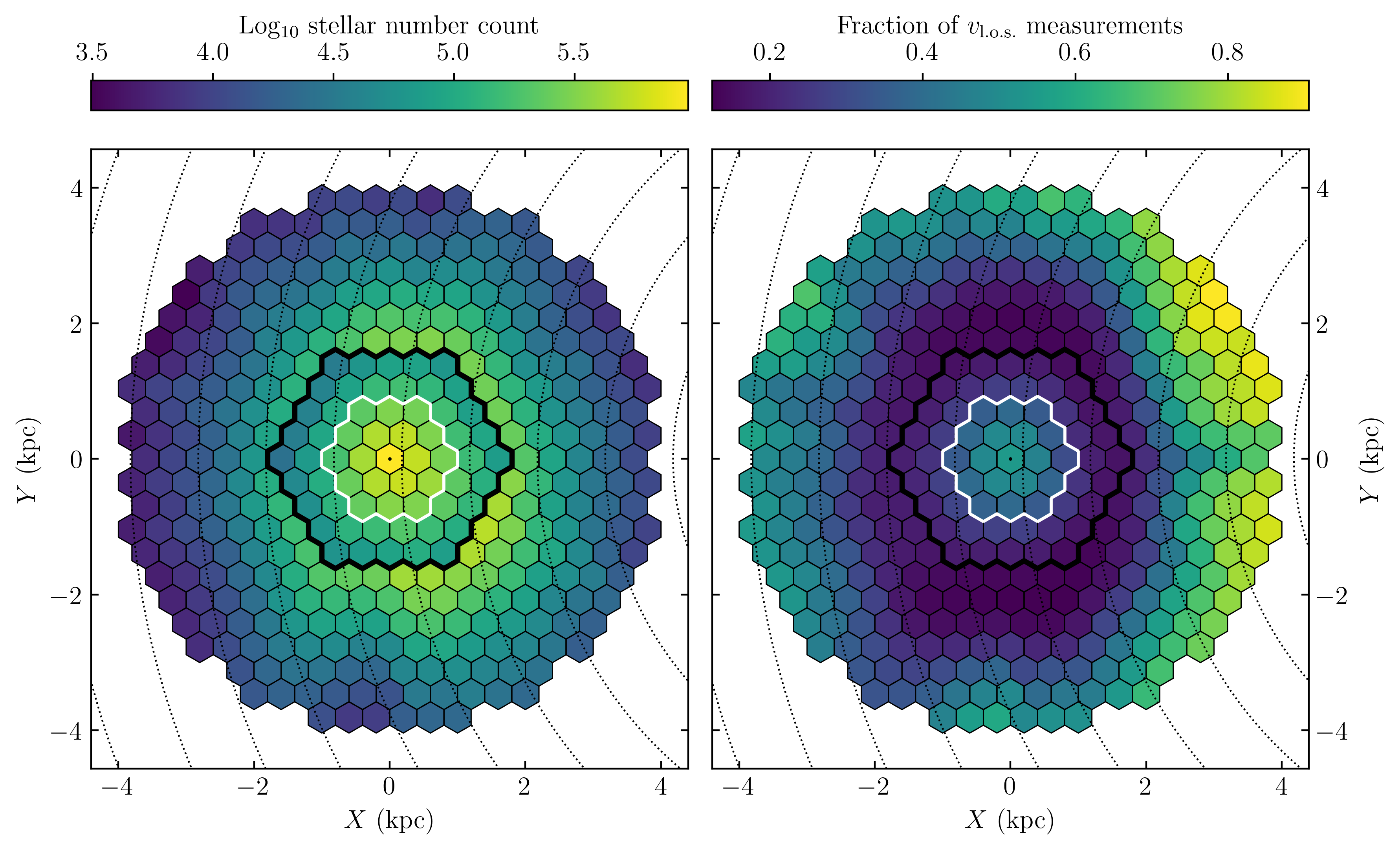}
    \caption{Stellar number counts for the spatially binned data samples in the Galactic plane. The left panel shows the number counts after data quality cuts and line-of-sight velocity cuts (see Section~\ref{sec:spatial_binning}). The right panel shows the fraction of stars, after data quality cuts, with \emph{Gaia} DR3 line-of-sight velocity measurements. The small black dot in the center shows the Sun's position. Dotted lines are contours of $R=i~\kpc$, where $i$ is an integer. The thick black line encloses the region where the line-of-sight velocity cut is applied. The inner white line encloses the region where we further subdivide the data in disk-plane parallel velocities (what we call the phase-space binning, described in detail in Section~\ref{sec:phase_space_binning}).}
    \label{fig:num_count}
\end{figure*}

In order to isolate a well defined phase spiral, we further restrict these data samples to stars that are on similar orbits, roughly following the bulk motion of the stellar disk. For the nearby data samples where $\vlos$ are required, we make a cut in angular momentum $L_z$: we first remove extreme $L_z$ outliers that are more than five standard deviations from the mean; we then recalculate those quantities and exclude stars that are beyond one standard deviation of the mean. For the distant data samples, we do the analogous cut but instead on the longitudinal velocity $v_l$, since full 3d velocity measurements are unavailable.

The number counts for the spatially binned data samples are shown in Figure~\ref{fig:num_count}. As one would expect, the fraction of stars with $\vlos$ measurements decreases with distance from the Sun, at least within roughly 2~kpc. Perhaps counter-intuitively, beyond this distance the fraction starts to increase again. This happens because in the most distant area cells only the brightest and easily observed stars will pass the quality cut on parallax precision, which are then also more likely to have $\vlos$ measurements.

\subsection{Phase-space binning}
\label{sec:phase_space_binning}

\begin{figure}
    \centering
    \includegraphics[width=1.0\columnwidth]{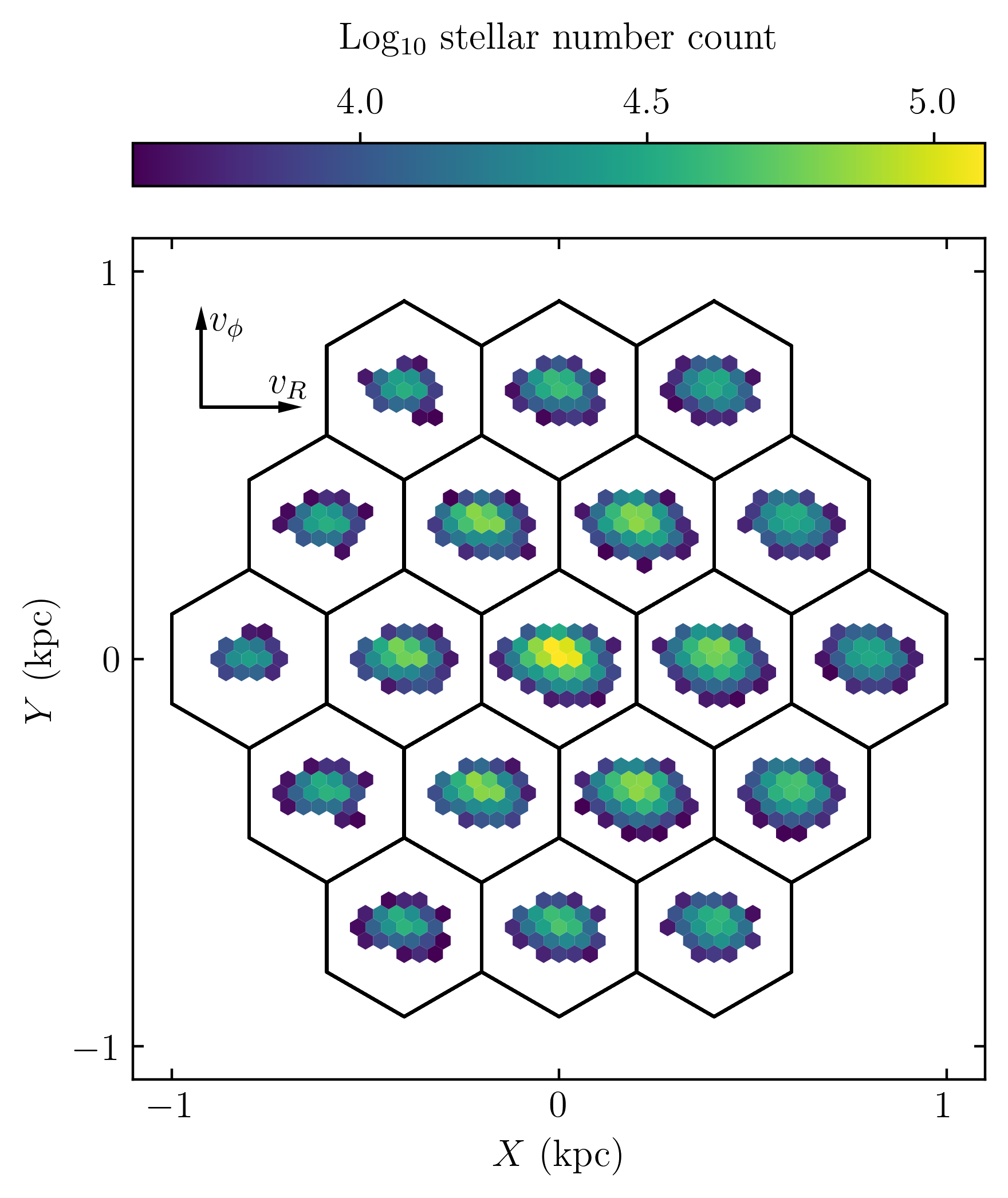}
    \caption{Stellar number counts for the phase-space binned data samples. The large hexagons drawn in black lines represent spatial area cells, with a grid spacing of $400~\pc$ (corresponding to the white outline in Figure~\ref{fig:num_count}). Within each large hexagon, the smaller colored hexagons represent further subdivision in $v_R$ and $v_\phi$ velocities, which have a grid spacing of $25~\kmsec$. The arrows in the top left show the directions of the two velocity components.}
    \label{fig:num_count_XYUV}
\end{figure}

For the spatial hexagon bins whose mid-points are within a distance of one kilo-parsec, we create another category of data sample, by further subdividing them in $v_R$ and $v_\phi$. We use the same quality cuts as outlined in Section~\ref{sec:quality_cuts}, and require an available \emph{Gaia} $\vlos$ measurement. For each separate spatial bin, we locate the mean $v_R$ and $v_\phi$ and take them to be the velocity grid's mid-point. We then tile the $(v_R,v_\phi)$ velocity plane using hexagons with a $25~\kmsec$ grid spacing. For each phase-space bin (i.e. divided in both space and velocity), we require a stellar number count above 4000, giving 508 separate data samples. This is illustrated in Figure~\ref{fig:num_count_XYUV}.

\section{Methods}
\label{sec:methods}

Our model of inference for fitting the vertical phase spiral closely follows that of \cite{Widmark-spiral-I,Widmark-spiral-II,Widmark-spiral-IV,Widmark-spiral-III}, but with two main modifications. Firstly, we handle the strong selection effects in a simpler, yet effective, manner (see Section~\ref{sec:2d_hist}). Secondly, the spiral is parametrized in terms of its present-day morphology, rather than with quantities that have a direct physical interpretation (see Section~\ref{sec:morph_params}). The free parameters of the spiral model are listed in Table~\ref{tab:model_parameters}. The gravitational potential, stellar orbits, and stellar density distributions are described in detail below.

{\renewcommand{\arraystretch}{1.6}
\begin{table}[ht]
	\centering
	\caption{Parameters in our model of inference. See Sections~\ref{sec:coords} and \ref{sec:morph_params} for precise definitions.}
	\label{tab:model_parameters}
    \begin{tabular}{| l | l |}
		\hline
            $\popp_\odot$  & Sun's phase-space coordinates \\
            \hline
            $Z_\odot$ & Sun's height \\
            $W_\odot$ & Sun's vertical velocity \\
		\hline
		\hline
		$\popp_\text{B}$  & Bulk density parameters \\
		\hline
		$a_k$ & Weights of Gaussian mixture model \\
		$\sigma_{z,k}$ & Spatial dispersions of Gaussian mixture model \\
            $\sigma_{w,k}$ & Velocity dispersions of Gaussian mixture model \\
            \hline
		\hline
		$\popp_\text{S}$  & Spiral model parameters \\
		\hline
		$A_\Phi$ & Gravitational potential scaling \\
		$\omega$ & Winding between heights 300--800~pc \\
		$\varphi_{600}$ & Present-day rotation phase for height 600~pc \\
		$\alpha$ & Single arm density amplitude \\
            $\beta$ & Double arm density amplitude \\
		\hline
	\end{tabular}
\end{table}}

\subsection{Gravitational potential and stellar orbits}
\label{sec:pot_and_orbits}

\subsubsection{Vertical dimension}
\label{sec:vertical_orbits}

The vertical gravitational potential is modeled as
\begin{equation}
\label{eq:phi_model}
    \Phi(z) = A_\Phi \, \Phi_\odot(z),
\end{equation}
where $A_\Phi$ is a scaling constant and $\Phi_\odot(z)$ is taken from a model of matter density components in the solar neighborhood. In our model of inference, $A_\Phi$ is free to vary while $\Phi_\odot(z)$ remains fixed.

Our solar neighborhood model is given by
\begin{equation}
\label{eq:solar_phi}
\begin{split}
    & \Phi_\odot(z) = 4 \pi G \times \bigg\{
    (0.01~\Msunppcc) \, \frac{z^2}{2} \\
    & + (120^2 \times 0.043~\Msun \pc^{-1}) \, \log \bigg[\text{cosh}\bigg(\frac{z}{120~\pc}\bigg)\bigg] \\ 
    & + (430^2 \times 0.045~\Msun \pc^{-1})\, \log \bigg[ \text{cosh}\bigg(\frac{z}{430~\pc}\bigg)\bigg]
    \bigg\}.
\end{split}
\end{equation}
This analytic function is fitted to the sum of matter density components in the solar neighborhood \citep{Schutz2017}, which is described in detail in Appendix~\ref{app:grav_pot}.

Under the assumption of vertical separability and a static gravitational potential, the vertical period of a star is given by
\begin{equation}\label{eq:period}
    P(E_z) = \oint \frac{\de z}{w} =  4\int_0^{z_\text{max}} \dfrac{\de z}{\sqrt{2[E_z-\Phi(z)]}},
\end{equation}
where
\begin{equation}\label{eq:Ez}
E_z = \Phi(z) + w^2 / 2
\end{equation}
is vertical energy per mass and $z_\text{max}$ is the orbit's maximum height.

We define a vertical phase angle, which describes where a star is in its vertical oscillation. This angle has an implicit dependence on $z$, $w$ and $\Phi(z)$, and is equal to
\begin{equation}\label{eq:phase_angle}
\begin{split}
    \theta_z = &
    \begin{cases}
    2 \pi P^{-1} Q & \text{if}\,z\geq0\,\text{and}\,w\geq0, \\
    \pi - 2 \pi P^{-1}Q & \text{if}\,z\geq0\,\text{and}\,w<0, \\
    \pi + 2 \pi P^{-1}Q & \text{if}\,z<0\,\text{and}\,w<0, \\
    2\pi - 2 \pi P^{-1}Q & \text{if}\,z<0\,\text{and}\,w\geq0.
    \end{cases}
\end{split}
\end{equation}
where
\begin{equation}
    Q = \displaystyle\int_0^{|z|}\dfrac{\de z}{\sqrt{2[E_z-\Phi(z)]}}.
\end{equation}

\subsubsection{Disk plane dimensions}
\label{sec:disk_plane_orbits}

In order to analyze our results, we also consider the respective data samples' motion within the disk plane, parameterized by either $(X,Y)$ or $(R,\phi)$. We model the in-plane orbits assuming an axisymmetric potential described by the rotational velocity curve
\begin{equation}\label{eq:rot_vel_curve}
    v_c(R) = v_{c,\odot} + s\times(R-R_\odot),
\end{equation}
where $v_{c,\odot} = 234~\kmsec$ is the rotational velocity at the solar radius, and $s = -2~\kmsec \,\pc^{-1}$ is its slope \citep{Zhou2023,Ou2024}.

Stellar orbits in the disk plane, with radius $R$ and angular momentum $L_z$, can be described by an effective gravitational acceleration \citep{BinneyAndTremaine2008}, according to
\begin{equation}
    F_\mathrm{eff.}(R, L_z) =
    \frac{L_z^2}{R^3} - \frac{v_c(R)^2}{R}.
\end{equation}
For perfectly circular orbits, the two terms on the right hand side cancel. A star's plane parallel position evolves according to these differential equations:
\begin{equation}
\begin{split}
    \frac{\de v_R}{\de t} & = F_\mathrm{eff.}(R, L_z), \\
    \frac{\de R}{\de t} & = v_R, \\
    \frac{\de \phi}{\de t} & = \frac{L_z}{R^2}.
\end{split}
\end{equation}

\subsection{Constructing vertical phase-space histograms}
\label{sec:2d_hist}

For each data sample, produced using either a spatial or phase-space binning, we summarize the data by constructing a 2d histogram in the vertical phase-space plane, written $N_{ij}$. The indices $i$ and $j$ loop over 120 bins in $z$ and 120 bins in $w$, in ranges $[-1,1]~\kpc$ and $[-3\sigma_w,3\sigma_w]$, where $\sigma_w$ is the data sample's vertical velocity dispersion.

Many data samples are affected by severe selection effects, mainly due to dust extinction and stellar crowding. These selection effects are difficult to model accurately, but due to our quality cuts (see Section~\ref{sec:data}) selection is largely only dependent on $z$, typically manifested as incompleteness bands close to the disk mid-plane. As shown in \cite{Widmark-spiral-IV,Widmark-spiral-III}, we can still precisely and robustly infer the vertical gravitational potential selection, as long as the phase spiral's shape can be extracted. To achieve this, we renormalize the vertical phase-space histogram by a factor $\eta_i$, where $i$ indexes bins in $z$. This renormalization factor fulfills that
\begin{equation}
    \eta_i \sum_j N_{ij}  \propto \mathrm{sech}^2\bigg( \frac{z_i}{300~\pc} \bigg),
\end{equation}
where we sum over $w$-bins at a given $z$-bin.
In other words, instead of modeling complicated selection effects, we simply impose a fixed shape for the stellar number density profile, described by a single $\mathrm{sech}^2$ with a scale height of 300~pc. Modifying this profile shape within reasonable limits (e.g. changing the scale height) has a negligible effect on our end results.

\subsection{Bulk background distribution}
\label{sec:bulk_model}

In steady state based methods, for example Jeans analysis, what we call the bulk background density is the key quantity that is used to infer the gravitational potential. In this method, the gravitational potential is instead inferred from the shape of the phase spiral. Here, the bulk is treated purely as a background and a nuisance distribution, fitted solely as a means to extract the phase spiral.

Our bulk background distribution is modeled with a mixture model of six bi-variate Gaussian distributions, indexed by $k=\{1,2,...,6\}$. It is equal to
\begin{equation}\label{eq:bulk_density}
\begin{split}
    & B(z,w\,|\,\popp_\mathrm{B}) = \\
    & \sum_{k=1}^{6} a_k \,
    \dfrac{\exp\Bigg(-\dfrac{z^2}{2\sigma_{z,k}^2}\Bigg)}{\sqrt{2\pi\sigma_{z,k}^2}} \,
    \dfrac{\exp\Bigg[-\dfrac{(W+W_\odot)^2}{2\sigma_{w,k}^2}\Bigg]}{\sqrt{2\pi\sigma_{w,k}^2}}.
\end{split}
\end{equation}
The six Gaussians are all centered on the origin of the $(z,w)$-plane, and have zero-valued $z$--$w$ correlation values, giving a bulk density that is mirror symmetric with respect to both $z$ and $w$. The free parameters of the bulk background distribution are amplitudes ($a_k$) and phase-space dispersions ($\sigma_{z,k}$ and $\sigma_{w,k}$), as listed in Table~\ref{tab:model_parameters}.

\subsection{Idealized spiral model}
\label{sec:spiral_model}
We model the phase spiral under the following simplifying assumptions: (i) the vertical dynamics are separable, such that motion parallel to the disk plane can be ignored; (ii) the perturbation that gives rise to the spiral does not have any winding to begin with; (iii) the gravitational potential is static and there are no self-gravity effects such as winding delay.

Because the disk's vertical gravitational potential is anharmonic, a non-equilibrium structure in the $(z,w)$-plane will wind into a spiral, since stars with higher vertical energies also have longer vertical periods. Winding is faster if the vertical gravitational potential is either steeper, corresponding to a heavier disk, or more anharmonic, corresponding to a more pinched (i.e. mid-plane concentrated) total matter density distribution. As described in Section~\ref{sec:vertical_orbits}, we only vary the vertical potential in terms of its amplitude. We do not include parameters to model the shape of $\Phi(z)$, since this cannot be robustly inferred. This shortcoming has been pointed out in previous work on weighing the Galactic disk using the phase spiral (e.g. \citealt{Widmark-spiral-I,Widmark-spiral-II}), and is the case generally for dynamical mass measurements. Even though an anharmonicity parameter is not directly included in our model of inference, it is still a crucial factor to consider when interpreting the results and relating the present-day morphological spiral parameters to physical quantities (see Section~\ref{sec:param_relations}).

We describe the relative density perturbation of the phase spiral as a function of vertical energy ($E_z$) and vertical phase-space angle ($\theta_z$, defined in Eq.~\ref{eq:phase_angle}). It is given by
\begin{equation}\label{eq:spiral_rel_density}
\begin{split}
    & S(E_z,\theta_z\,|\,\popp_\text{spiral}) = \\
    & \alpha \cos\big[ \theta_z-\varphi(E_z) \big] + \\
    & \beta \cos\big[ 2\theta_z-2\varphi(E_z) \big],
\end{split}
\end{equation}
where $\alpha$ and $\beta$ are the anti-symmetric and symmetric relative density amplitudes, corresponding to single and double armed spirals.
The quantity
\begin{equation}
\label{eq:rotation_of_Ez}
    \varphi(E_z) = \varphi_\mathrm{init.} + \frac{2 \pi t_\omega}{P(E_z)},
\end{equation}
traces the peak of the spiral over-density in vertical phase angle, which depends on the perturbation's initial angle ($\varphi_\mathrm{init.}$) and the winding time ($t_\omega$). Thus the amount of winding between two vertical energies is equal to
\begin{equation}
\label{eq:winding_of_Ez}
\begin{split}
    \omega(E_{z,1},E_{z,2}) & = \varphi(E_{z,1})-\varphi(E_{z,2}) \\
    & = 2 \pi t_\omega \times \big[ P(E_{z,1})^{-1}-P(E_{z,2})^{-1}\big],
\end{split}
\end{equation}
given in radians, which is positive if $E_{z,1}<E_{z,2}$.

We do not want to model the spiral over the full $(z,w)$-plane. For this reason we define a mask function equal to
\begin{equation}
\label{eq:mask}
\begin{split}
    M(E_z) = \; & \mathrm{sigm}\bigg\{ \frac{E_z - \Phi(300~\pc)}{(25~\pc) \times  [\de \Phi / \de z]_{z=300~\pc}} \bigg\} \times \\
    & \mathrm{sigm}\bigg\{ \frac{\Phi(800~\pc)-E_z}{(25~\pc) \times  [\de \Phi / \de z]_{z=800~\pc}} \bigg\},
\end{split}
\end{equation}
where $\mathrm{sigm}(x) = 1 / [1+\exp(-x)]$ is a sigmoid function that transitions smoothly from zero to unity when its argument goes from negative to positive. In the mask function, the two sigmoid factors represent lower and upper limits in vertical energy, corresponding to $\Phi(300~\pc)$ and $\Phi(800~\pc)$.

\subsection{Spiral parametrization}\label{sec:morph_params}

\begin{figure*}
    \centering
    \includegraphics[width=0.65\textwidth]{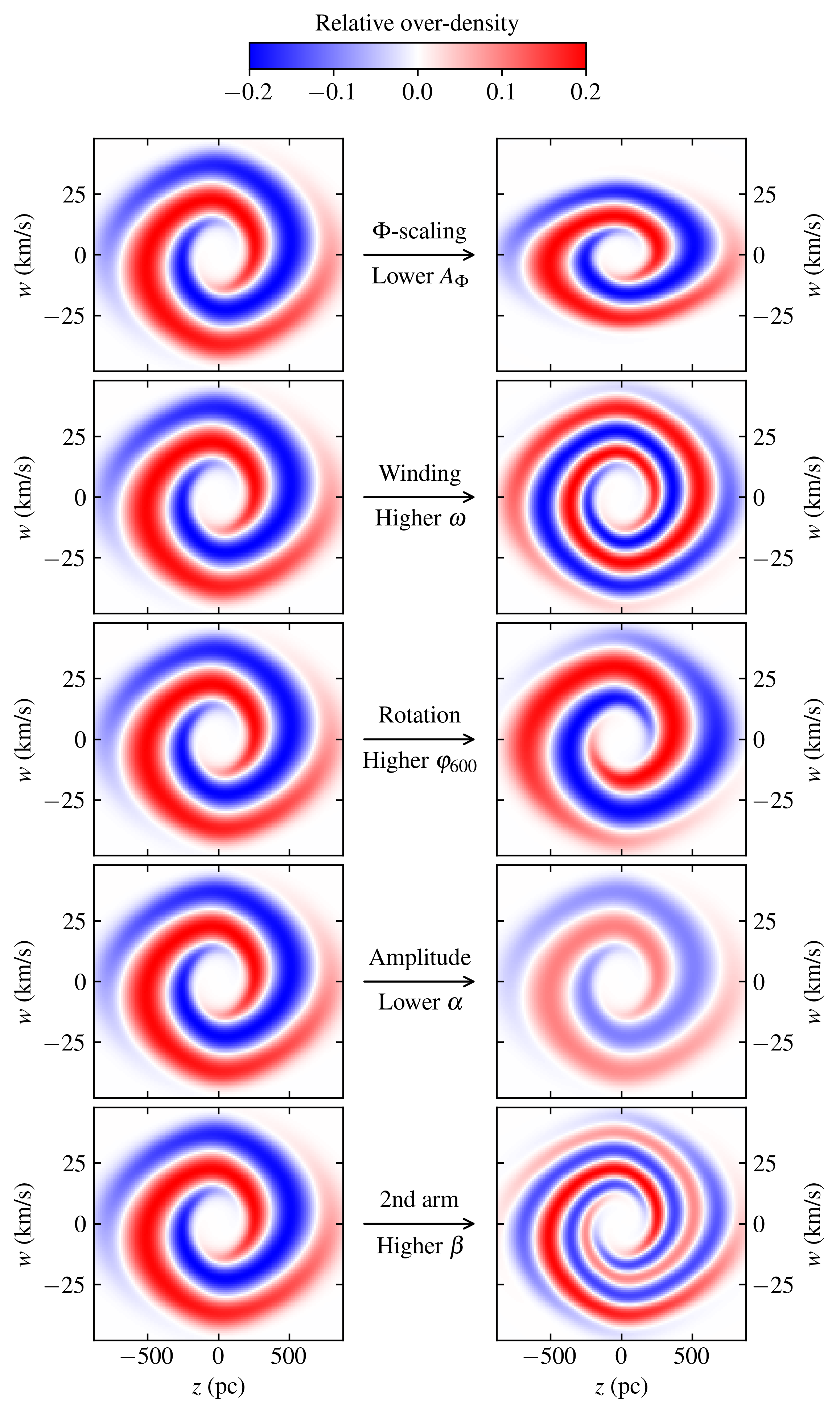}
    \caption{Schematic of spiral parameters: $A_\Phi$, $\omega$, $\varphi$, $\alpha$, $\beta$. This figure has a pedagogical purpose, where each row illustrates how the spiral changes when varying only one of the morphological parameters.}
    \label{fig:spiral_schematic}
\end{figure*}

Figure~\ref{fig:spiral_schematic} illustrates the five free parameters of the phase spiral in our model of inference: gravitational potential scaling ($A_\Phi$), winding ($\omega$), phase angle ($\varphi_{600}$), amplitudes of single armed ($\alpha$) and double armed ($\beta$) components of the spiral density perturbation. These parameters, also listed in Table~\ref{tab:model_parameters}, have a direct phenomenological meaning in terms of the spiral's present-day morphology. This parametrization avoids strong degeneracies that arise between parameters with a more direct physical interpretation (e.g. between winding time and gravitational potential). The morphological parameters are of course related to physical quantities, which is discussed in Section~\ref{sec:param_relations} below.

Winding is parametrized by the difference in the spiral's vertical phase angle between anchor heights of $z_A=300~\pc$ and $z_B=800~\pc$. Thus it is equal to $\omega[\Phi(300~\pc), \Phi(800~\pc)]$, using the definition of winding from Eq.~\eqref{eq:winding_of_Ez}. Henceforth in this article, we write only $\omega$ for shorthand.

The spiral's rotation is parametrized by the rotation phase of the spiral over-density at an anchor height of $600~\pc$, given by $\varphi[\Phi(600~\pc)]$ using the definition from Eq.~\eqref{eq:rotation_of_Ez}. Henceforth, we write $\varphi_{600}$ for shorthand. Given that the shape of $\Phi(z)$ is fixed in our model, the phase angle of a different height is given by a linear combination with $\omega$: for example, $\varphi_{300} = \varphi_{600} - 0.624\,\omega$, where the numerical value depends on the precise shape of $\Phi_\odot$.

\subsection{Fitting procedure}

The fitting procedure is applied separately and independently to each data sample. We first fit the bulk background distribution to the renormalized 2d histogram ($\eta_i N_{ij}$), in the absence of any phase spiral perturbation. For the phase-space binned data samples, we fix $W_\odot = 7.25~\kmsec$, while for the spatially binned data samples $W_\odot$ is free to vary and fitted jointly with $\popp_\mathrm{B}$. In all fits, we fix $Z_\odot = 20~\pc$.\footnote{Ideally, we would let $Z_\odot$ be a free parameter, since the Galactic disk plane is not perfectly flat. However, we cannot robustly infer $Z_\odot$ due to spatially dependent selection effects. We expect that an incorrectly assumed $Z_\odot$ will give rise to a systematic bias which is most severe for $A_\Phi$, while $\omega$ and $\varphi_{600}$ should still be robust. We refer to \cite{Widmark-spiral-IV} for more information, where they performed tests on simulations to quantify this bias.}

We maximize the likelihood
\begin{equation}
\label{eq:bulk_likelihood}
    \log\mathcal{L} = -\sum_{ij} \log \Bigg\{ \mathrm{cosh}\Bigg[
    \frac{B(z_i,w_j)-\eta_i N_{ij}}{\sqrt{\eta_i B(z_i,w_j)}}
    \Bigg] \Bigg\} + \mathrm{const.},
\end{equation}
summing over all bins of the 2d histogram.
We use this functional form, instead of a standard squared error, in order to be less penalizing towards strong outlier values. The function $\log[\mathrm{cosh}(x)]$ has a quadratic shape for small argument values (i.e. $x\lesssim 1$), but approaches a linear asymptote for large argument values. We include the renormalization factor $\eta_i$ in the argument's denominator, effectively lowering the weight for histogram bins with low completeness.

In the second step of our fitting procedure, we fit the relative density perturbation of the phase spiral, with $\popp_\odot$ and $\popp_\mathrm{B}$ fixed, while $\popp_\mathrm{S}$ is free to vary. We maximize the likelihood
\begin{equation}
    \log\mathcal{L} = -\sum_{ij} \log \Bigg\{ \mathrm{cosh}\Bigg[
    \frac{f(z_i,w_j)-\eta_i N_{ij}}{\sqrt{\eta_i f(z_i,w_j)}}
    \Bigg] \Bigg\} + \mathrm{const.},
\end{equation}
where
\begin{equation}\label{eq:total_density}
    f(z,w) = B(z,w) \Big[ 1 + M(E_z)\, S(E_z,\theta_z) \Big]
\end{equation}
is the total stellar number density of our model (where model parameter dependencies are omitted for shorthand; see above equations for details).

\section{Results}
\label{sec:results}

As described in Section~\ref{sec:data}, we construct several hundred stellar samples, using cuts in disk-plane parallel phase-space coordinates ($X$, $Y$, $v_R$, $v_\phi$). There are two main categories of data samples, where the first category is further subdivided in two:
\begin{outline}
    \1 Spatially binned
        \2 Nearby ($\leq 1.6~\kpc$, $\vlos$ required)
        \2 Distant ($> 1.6~\kpc$, $\vlos$ not required)
    \1 Phase-space binned ($<1~\kpc$, $\vlos$ required)
\end{outline}

For each data sample, we independently fit a phase spiral as a relative stellar density perturbation in the $(z,w)$-plane. All individual data samples and their respective fitted phase spirals have been studied by eye. Data samples that did not produce a convincing fit are omitted.

Some supplementary results and figures are shown in Appendix~\ref{app:supp_results}. They are complementary to the results presented in this section.

\subsection{Spatially binned data samples}
\label{sec:results_XY}

\begin{figure*}
    \centering
    \includegraphics[width=1.0\textwidth]{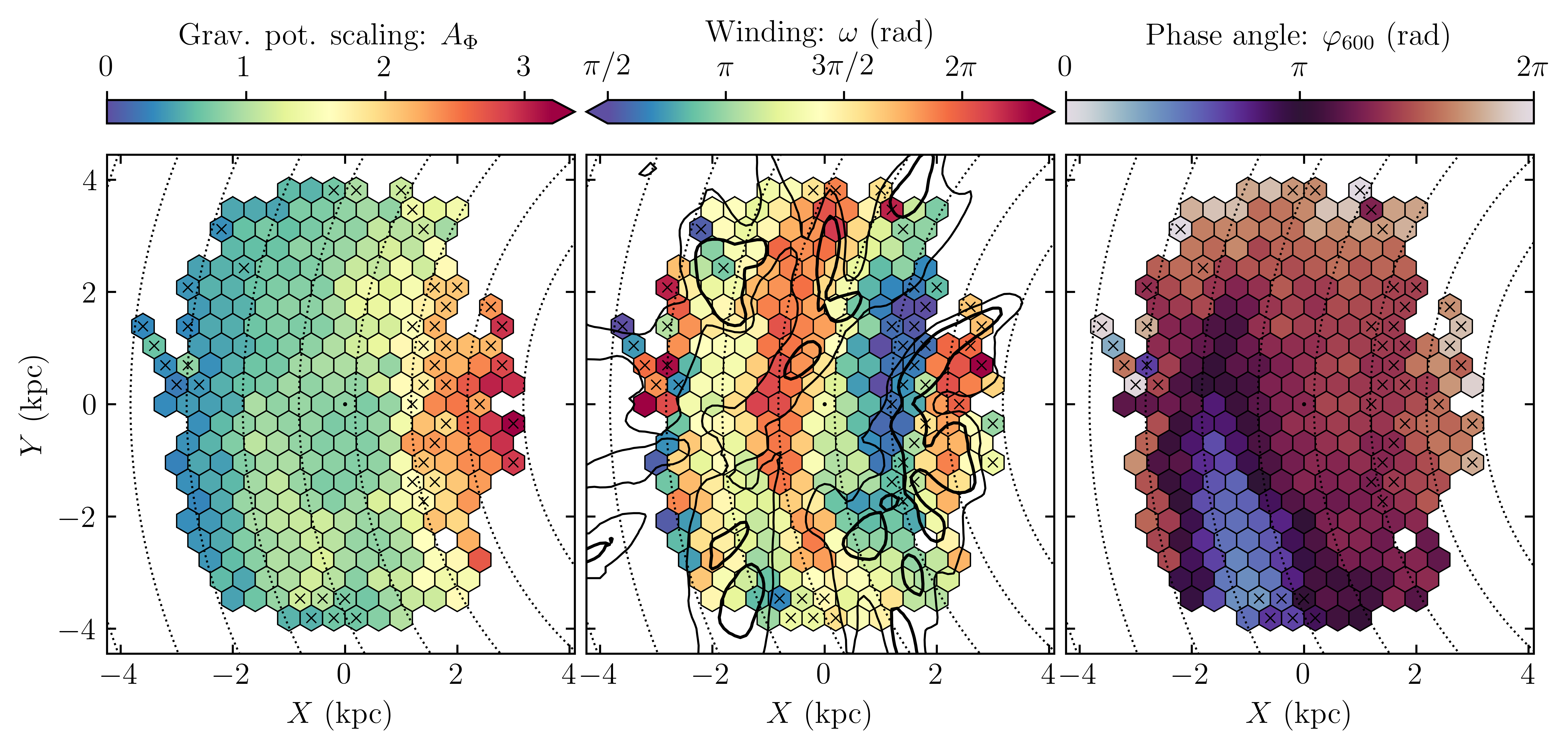}
    \caption{Spiral parameters ($A_\Phi$, $\omega$, $\varphi_{600}$) for our spatially binned data samples in the $(X,Y)$-plane. The right panel, showing the phase $\varphi_{600}$, has a cyclical color map. The Galactic center is towards the right, and the direction of Galactic rotation is upwards. In each panel, the black dot in the center shows the solar position, while the dotted lines are of constant $R$, as in Figure~\ref{fig:num_count}. All fitted spirals have been studied by eye; data samples without a visible phase spiral are omitted, and dubious fits are marked with a cross. In the middle panel, the overlaid black contours enclose regions with a high relative ratio of upper main sequence stars \citep[][see Appendix~\ref{app:supp_results} for details]{Poggio2021}. Such stars trace the Milky Way spiral arms; for example, the high valued region at $R\simeq 9~\kpc$ is associated with the Local Arm.}
    \label{fig:spat_plane_triple}
\end{figure*}

In Figure~\ref{fig:spat_plane_triple}, we show the inferred results for the spatially binned data samples (combined nearby and distant) in the Milky Way disk plane, for the vertical gravitational potential scaling ($A_\Phi$), winding ($\omega$), and rotation phase ($\varphi_{600}$). Data samples where the results were identified as dubious in our by-eye inspection are indicated with a cross.

\begin{figure}
    \centering
    \includegraphics[width=1.0\columnwidth]{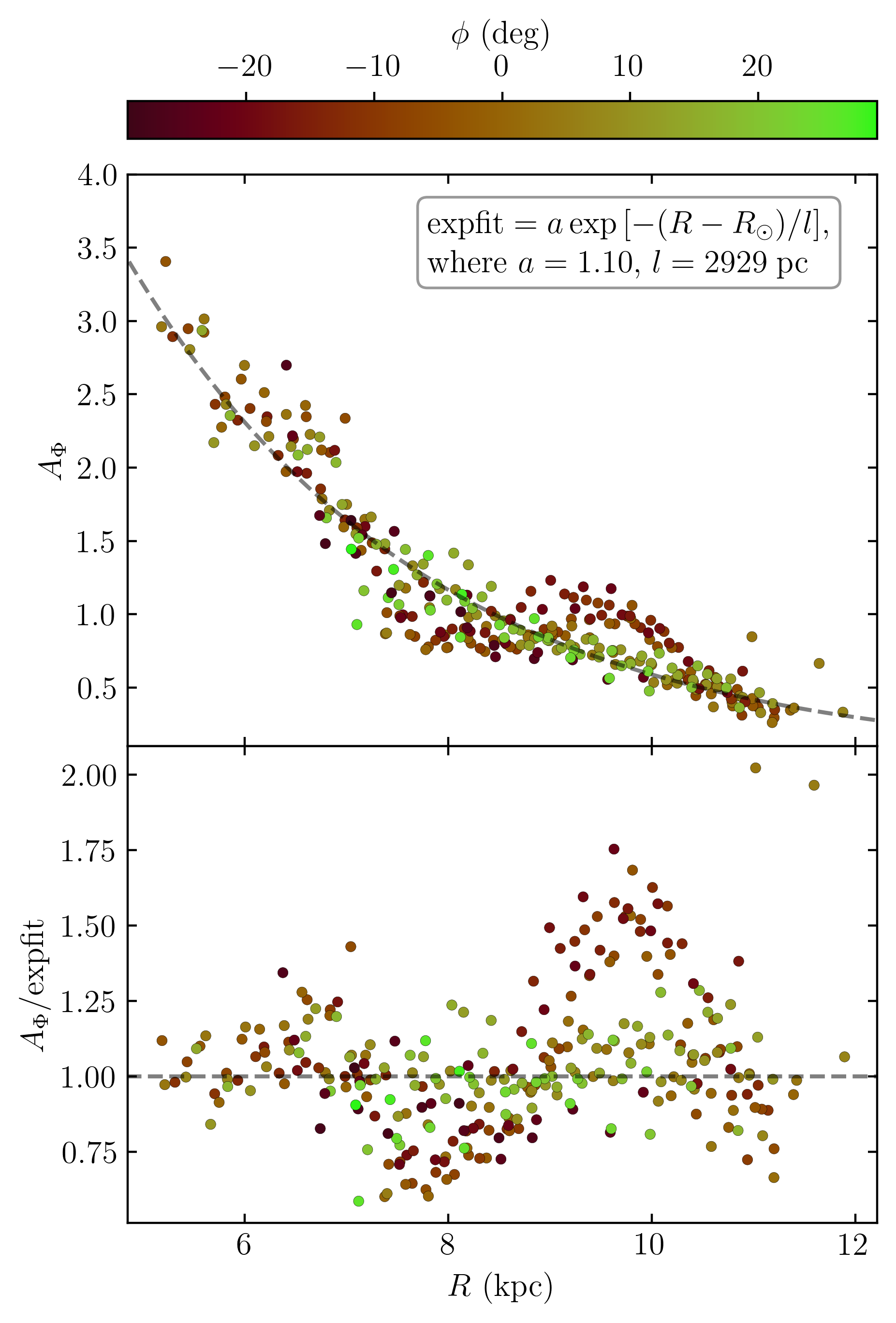}
    \caption{Gravitational potential scaling ($A_\Phi$) for the spatially binned data samples. The scatter points are colored by azimuth ($\phi$). The dashed line, seen in all panels, is an exponential function fitted to the results of the spatially binned data samples; see the main text for details.}
    \label{fig:scatter_potscaling_XY}
\end{figure}

The results for $A_\Phi$ are shown in the left panel of Figure~\ref{fig:spat_plane_triple}. We see a clear trend with Galactocentric radius, where low $R$ corresponds to larger $A_\Phi$.  This profile is more clearly illustrated in Figure~\ref{fig:scatter_potscaling_XY}, where we show the same relationship as a scatter plot with respect to $R$. Since $A_\Phi$ is a scaling parameter for the vertical gravitational potential (see Section~\ref{sec:vertical_orbits} and Appendix~\ref{app:grav_pot}), it is a close proxy for the thin disk surface density. In order to derive the thin disk scale length, we fit an exponentially decaying function of $A_\Phi$ with respect to $R$. In this fit, we use the L1-norm (in order to lessen the influence of strong outliers) and an uncertainty of $\sigma \propto \sqrt{N}$, where $N$ is a data sample's stellar number count. We derive a thin disk scale length of $2.9~\kpc$ and a solar radius value for $A_\Phi$ close to unity (i.e. in good agreement with our solar neighborhood matter density model). There are some spatially coherent deviations from the exponential fit, for example in the form of an over-dense region at approximately $(X,Y)=(-1,-2)~\kpc$. Perhaps this could be related to true variations in the disk surface density, but we refrain from drawing any strong conclusions from such structures, since $A_\Phi$ is probably particularly sensitive to systematic biases in distant disk regions, for example due to selection effects and deviations from the assumed mid-plane height.

The results for $\omega$, shown in the middle panel of Figure~\ref{fig:spat_plane_triple}, has clear structure on smaller ($\lesssim 1~\kpc$) spatial scales. There are bands of high winding, most prominently at $R\simeq 9~\kpc$ and also at $R\simeq 6~\kpc$, which correlate with the locations of the Local Arm and Sagittarius-Carina Arm (\citealt{Poggio2021,Poggio2022}; see Appendix~\ref{app:supp_results} and Figure~\ref{fig:poggio}). Curiously, apart from these smaller-scale variations between roughly $\pi/2$ and $2\pi$, $\omega$ does not have a significant overall slope with respect to $R$.

The results for $\varphi_{600}$, shown in the right panel of Figure~\ref{fig:spat_plane_triple}, is remarkably smooth and uniform over large spatial scales. There is a weak slope, whereby $\varphi_{600}$ increases towards the top right (i.e. towards positive $X+Y$), with variations on the order of $3 \pi / 2$ over the complete observed spatial area. There seems to be some smaller-scale structure in the lower left corner, for $Y<0$ and $R \gtrsim 9~\kpc$, where $\varphi_{600}$ takes values close to $\pi / 2$.

We show the corresponding amplitude parameters ($\alpha$ and $\beta$) in Appendix~\ref{app:supp_results}. They are not included in the main text because they are likely significantly biased in a manner that depends strongly on distance as well as small-scale spatial selection effects. Spiral amplitude measurements are likely robust for spatially nearby data samples, where $\vlos$ measurements are required, but at greater distances selection effects become more severe and data uncertainties more significant. As long as the velocities are not biased, the shape of the phase spiral will still be robustly inferred. However, large uncertainties will effectively smear the spiral in the $(z,w)$-plane, thus inducing a smaller contrast in the stellar number density and biasing the amplitudes towards lower values.

\subsection{Phase-space binned data samples}
\label{sec:results_XYUV}

\begin{figure*}
    \centering
    \includegraphics[width=1.0\textwidth]{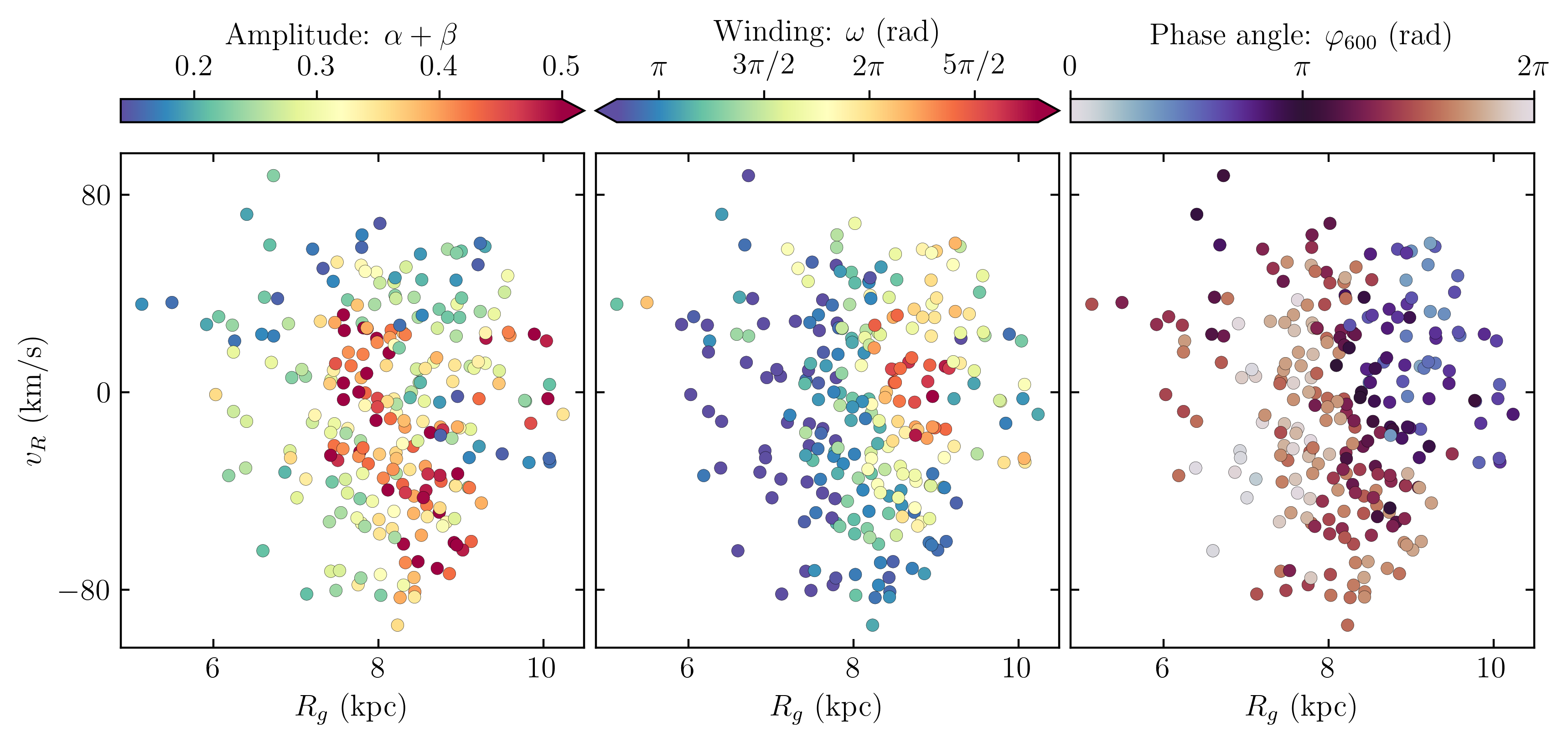}
    \caption{Spiral parameters ($\alpha + \beta$, $\omega$, $\varphi_{600}$) for the phase-space binning, in the plane of $R_g$ and $v_R$. We note that the color bar scale for $\omega$ is different from that of Figure~\ref{fig:spat_plane_triple}. For better visibility, a small Gaussian noise is added to both $R_g$ and $v_R$, with dispersions of $20~\pc$ and $4~\kmsec$; without this noise, there would be much overlap between nearby scatter points. All fitted spirals have been studied by eye, omitting poor fits. We also require an amplitude of $\alpha + \beta \geq 0.14$ for these results.}
    \label{fig:XYUV_triple}
\end{figure*}

In Figure~\ref{fig:XYUV_triple}, we show the inferred results for the phase-space binned data samples, for the total spiral amplitude ($\alpha+\beta$), winding ($\omega$), and phase ($\varphi_{600}$). Since the phase-space binning is made in $\{X,Y,v_\phi,v_R\}$, we must show our results in some projection of that four-dimensional space. Firstly, the natural choice is to show the phase spiral's dependence with respect to the guiding radius ($R_g$). Secondly, we found significant structure in terms of the data samples' epicyclic motion. This can be illustrated in various ways, for example in terms of epicyclic action ($J_R$) or action angle ($\theta_R$). However, any single parameter cannot encapsulate the full epicyclic information. By testing several options, we found the plane of guiding radius ($R_g$) and present-day Galactic radial velocity ($v_R$) to be the most illustrative, in the sense that structure is most clearly apparent. The parameter $v_R$ can be seen as a middle ground between $J_R$ and $\theta_R$: a high $|v_R|$ indicates high $J_R$, while the sign of $v_R$ correlates with $\theta_R$.

The results for $\alpha+\beta$ are shown in the left panel of Figure~\ref{fig:XYUV_triple}. Unlike the amplitudes for the spatially binned data samples, we consider the amplitudes for the phase-space binning to be robustly inferred, since these data samples are constructed from a nearby spatial volume and restricted to stars with $\vlos$ measurements. For some data samples, the amplitude is remarkably high, with values over $50~\%$. There is a ridge-like structure of high valued amplitudes, ranging from $(R_g,v_R)\simeq(8~\kpc,\,0~\kmsec)$ towards $(R_g,v_R)\simeq(9~\kpc,\,-80~\kmsec)$. The data samples around $(R_g,v_R)\simeq(8~\kpc,\,0~\kmsec)$ are on close-to-circular orbits, since neither the radial displacement ($R\simeq R_g \simeq R_\odot$) nor the radial velocity ($v_R \simeq 0~\kmsec$) contributes a high $J_R$.

When interpreting these results, it is important to consider the selection effects that are inherent to the data sample construction. Because the phase-space binned data samples are constructed from a spatially local sample, within $1~\kpc$, the only circular orbits we can observe have $R\simeq R_\odot$. Conversely, data samples where $R_g \gg R_\odot$ or $R_g \ll R_\odot$ are by necessity on eccentric orbits. Seeing high amplitudes for $R_g \simeq R_\odot$ does not necessarily mean that spirals are more pronounced at that guiding radii; it could equally well be that high amplitudes are correlated with circular orbits in general.

The results for $\omega$ are shown in the middle panel of Figure~\ref{fig:XYUV_triple}. There is a clear cluster of high winding, which differs from the regions of high amplitude. High winding is seen at $R_g \simeq 9~\kpc$, in particular for $v_R \simeq 0~\kmsec$. This is consistent with the results of the spatially binned data samples, where a band of $R\simeq 9~\kpc$ exhibits high winding. We see low $\omega$ values for low $R_g$ (with a few exceptions, e.g. for $R_g<6~\kpc$), which is largely consistent with the smaller scale variations we saw for the spatially binned data samples.

The results for $\varphi_{600}$ are shown in the right panel of Figure~\ref{fig:XYUV_triple}. Similar to the results of the spatially binned data samples, $\varphi_{600}$ varies smoothly and is close to uniform. There is a slope in the form of a negative correlation between $R_g$ and $\varphi_{600}$, which is fairly consistent with the slope seen in the spatially binned data samples. There is also a slight negative correlation between $\varphi_{600}$ and $v_R$.

\begin{figure*}
    \centering
    \includegraphics[width=1.0\textwidth]{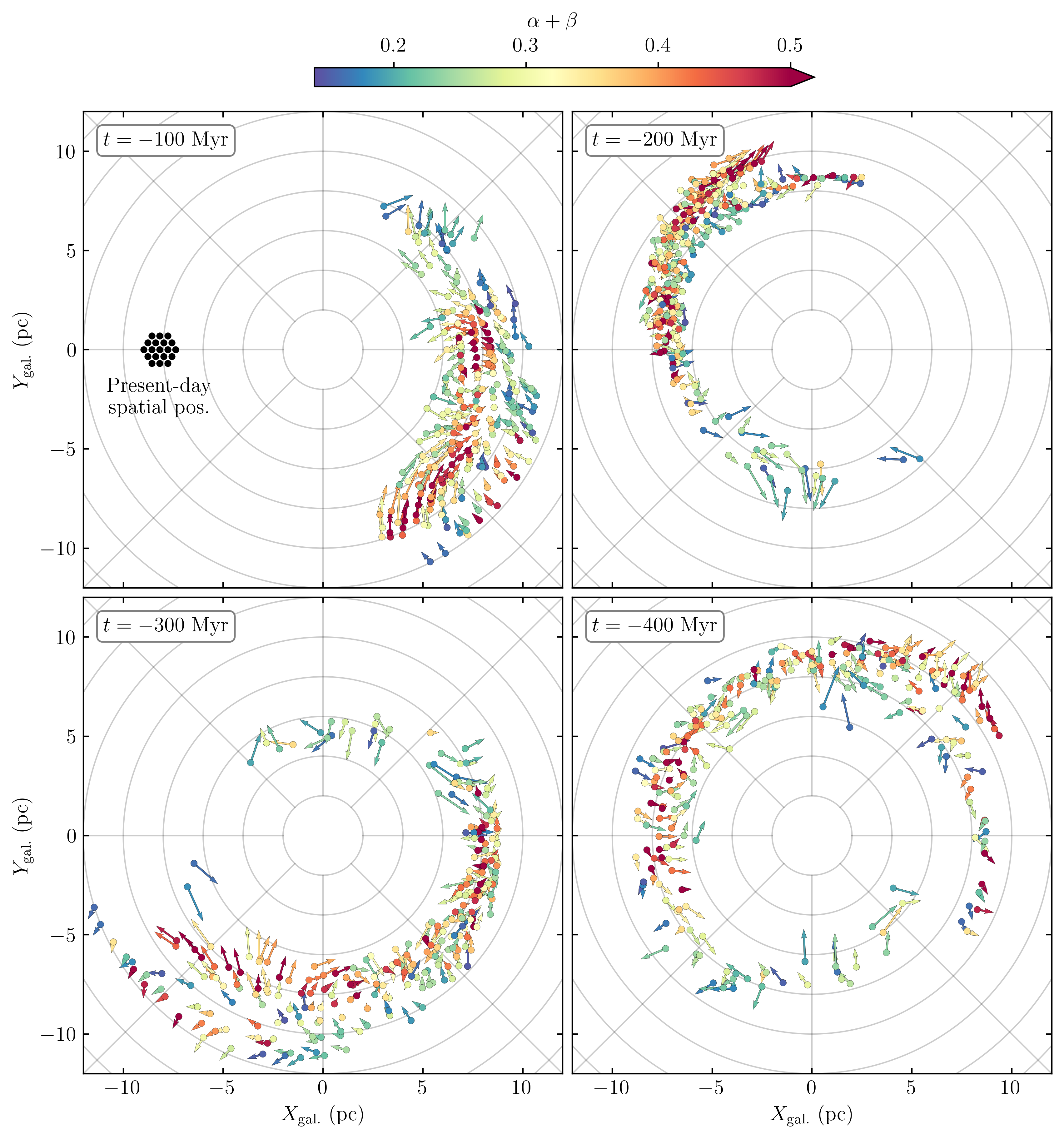}
    \caption{Total spiral amplitude ($\alpha + \beta$) for the phase-space binned data samples, in their back-propagated plane-parallel disk locations at time snapshots of 100, 200, 300, and 400~Myr ago (see the main text for further details). The arrows show disk parallel velocities relative to the velocity of a circular orbit at the respective data samples' momentary spatial positions; the arrow lengths are normalized to the square root of the disk parallel speed. The gray lines show constant Galactocentric radii and azimuth, with line spacings of 2~kpc and $45^\circ$.}
    \label{fig:rewind_amplitude}
\end{figure*}

In Figure~\ref{fig:rewind_amplitude}, we show the total spiral amplitude of our phase-space binned data samples, in terms of their back-propagated $(X,Y)$-coordinates, for four different time snapshots ($t=\{-100,-200,-300,-400\}~\Myr$). This highlights how the respective data samples wrapped around the Milky Way at the time that the spiral perturbation was produced (a few 100~Myr ago). Furthermore, it is clear that data samples that have been spatially close to each other, and have similar $R_g$, can still have quite different spiral amplitudes, with a strong dependence on epicyclic motion. If we focus on the data samples with high $R_g$ (i.e. the tail in the clockwise direction), then the high amplitude data samples can be temporarily on the inside (lower $R$, e.g. at $t=-300~\Myr$), or temporarily on the outside (higher $R$, e.g. at $t=-400~\Myr$), of lower amplitude data samples.

\subsection{Physical interpretation}
\label{sec:param_relations}

The spiral properties that we infer have a non-trivial and largely non-separable dependence on the four-dimensional disk plane parallel coordinates. The inferred spiral properties represent the observed present-day spiral morphology, which is related to physical quantities. However, these relationships are in reality more complicated than for the idealized model outlined in Section~\ref{sec:spiral_model}.

In our model of inference, the vertical gravitational potential has a fixed shape, since we cannot directly infer shape information with accuracy. In actuality, the vertical gravitational potential's level of harmonicity could differ between disk regions, which would affect the winding speed. A more anharmonic potential, corresponding to a more pinched (i.e. mid-plane concentrated) matter density distribution, would boost winding because of a higher vertical period difference between the two anchor heights. We parametrize the potential's anharmonicity using the quantity
\begin{equation}
    \Gamma = \frac{P[\Phi(z_A)]^{-1}-P[\Phi(z_B)]^{-1}}{P[\Phi(z_B)]^{-1}}.
\end{equation}
For the solar neighborhood model of the vertical potential, $P[\Phi(300~\pc)]=99.7~\Myr$ and $P[\Phi(800~\pc)]=125.7~\Myr$, giving $\Gamma = 0.260$. We refer to Appendix~\ref{app:grav_pot} for further details.

If we include effects of varying anharmonicity and self-gravity delay, the winding parameter is proportional to
\begin{equation}
\label{eq:winding_morph2phys}
    \omega \propto A_\Phi \Gamma \,(t_\text{pert.}-t_\text{delay}),
\end{equation}
where the times $t_\text{pert.}$ and $t_\text{delay}$ are perturbation and delay time, respectively.

We expect $\varphi_{600}$ to be largely independent of $\Gamma$. While $\omega$ depends on a vertical period difference, $\varphi_{600}$ depends on the period directly, such that $\Gamma$ is approximately negligible. Self-gravity induced delays could also have different effects on $\varphi_{600}$ and $\omega$. If we consider dynamics only in the disk's vertical dimension, then a winding delay would manifest itself as a dipole (or quadrupole) that spins like a bar in the $(z,w)$-plane.\footnote{There is another type of winding delay due to disk self-gravity in the form of ``stationary spirals'' \citep{Widrow2023}, which could arise due to swing amplification in shearing disks when excited by a massive cloud. However, such structures are local and spatially small phenomena.} This would delay winding, but $\varphi_{600}$ would still evolve in time, in largely the same manner. As a result, we would expect the following approximate relation:
\begin{equation}
\label{eq:phase_angle_morph2phys}
    \varphi_{600} = \varphi_\mathrm{init.} + \frac{2 \pi t_\text{pert.}}{P(600~\pc)}, \; \text{mod}\, 2\pi.
\end{equation}
It is of course possible, perhaps even likely, that winding delays are caused by more complicated, three-dimensional dynamics. For example, a passing satellite could give rise to some intermediate dynamical feature, such as a dark matter halo wake or corrugation modes propagating through the disk, that only later induces a $(z,w)$-dipole that winds into a spiral. In such a context, Eq.~\eqref{eq:phase_angle_morph2phys} could still be valid, if we consider $t_\text{pert.}$ to be given by the moment that the vertical phase-space dipole is induced.

\begin{figure}
    \centering
    \includegraphics[width=.84\columnwidth]{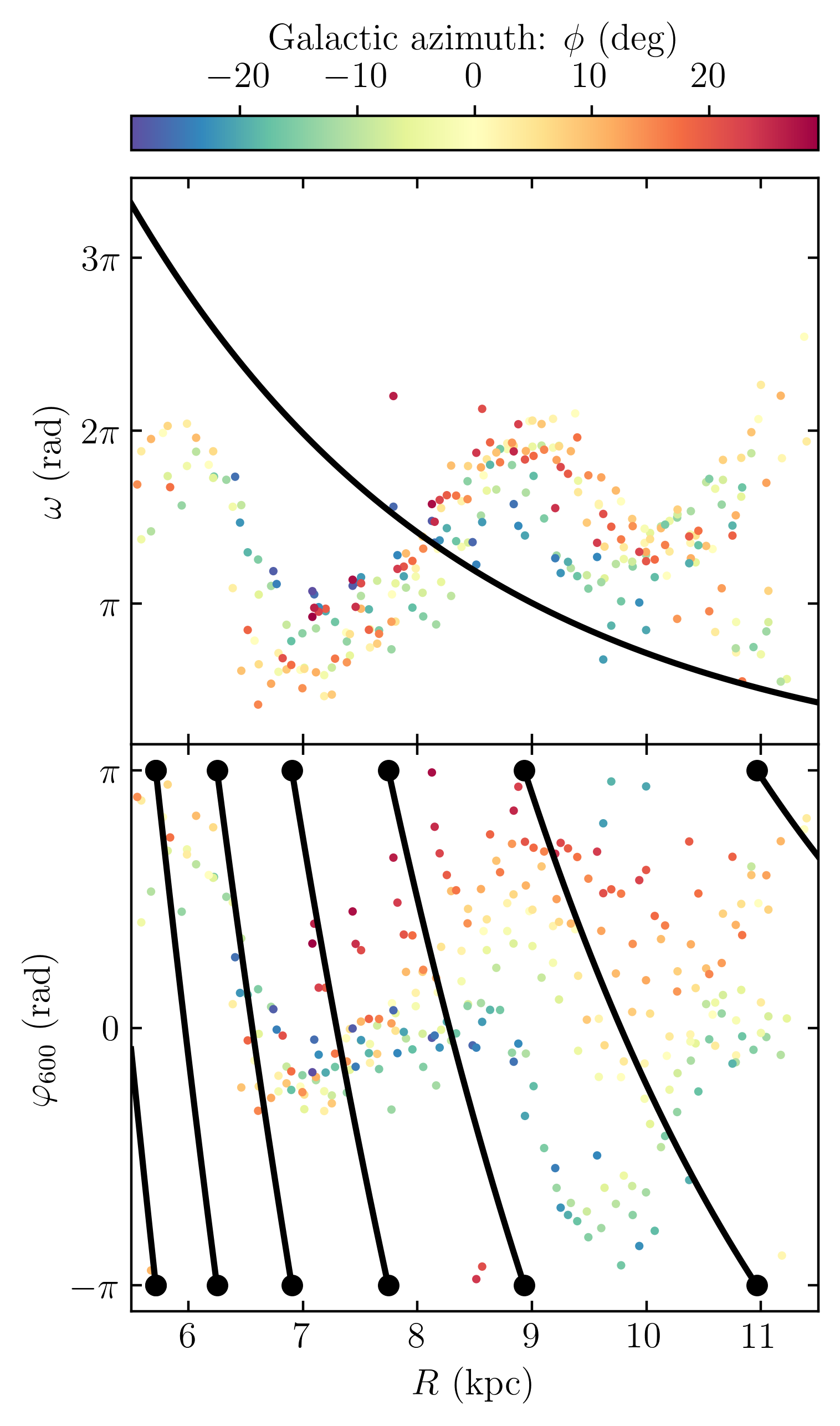}
    \caption{Black lines show radial profiles for winding ($\omega$, upper panel) and rotation phase ($\varphi_{600}$, lower panel), for an idealized model where we assume a uniform perturbation time of 300~Myr, an exponentially decaying disk surface density, and no self-gravity effects. In the bottom panel, the black circular markers denote artificial discontinuities due to the mod $2\pi$ operator. Scatter points show the measured parameter values for the spatially binned data samples, colored by Galactic azimuth. The idealized model has significant slopes for both $\omega$ and $\varphi_{600}$, which are not present in our results.}
    \label{fig:winding_rot_toy_model}
\end{figure}

In Figure~\ref{fig:winding_rot_toy_model}, we show radial profiles for $\omega$ and $\varphi_{600}$ in an idealized model, where we assume a spatially uniform perturbation time of $t_\text{pert.}=300~\Myr$, no self-gravity delay, and a surface density that decays exponentially with a scale length of 2.9~kpc. We clearly see very steep slopes with respect to $R$, where $\omega$ in the inner disk ($R\simeq6~\kpc$) is more than six times higher than that of the outer disk ($R\simeq11~\kpc$). Analogously, $\varphi_{600}$ in the inner and outer disk differs by several complete rotations, creating a sawtooth pattern through the mod $2\pi$ operator. This pattern is present also in test-particle simulations, as shown in Appendix~\ref{app:sim_results}.

\subsection{Comparison to data}

\begin{figure*}
    \centering
    \includegraphics[width=1.0\textwidth]{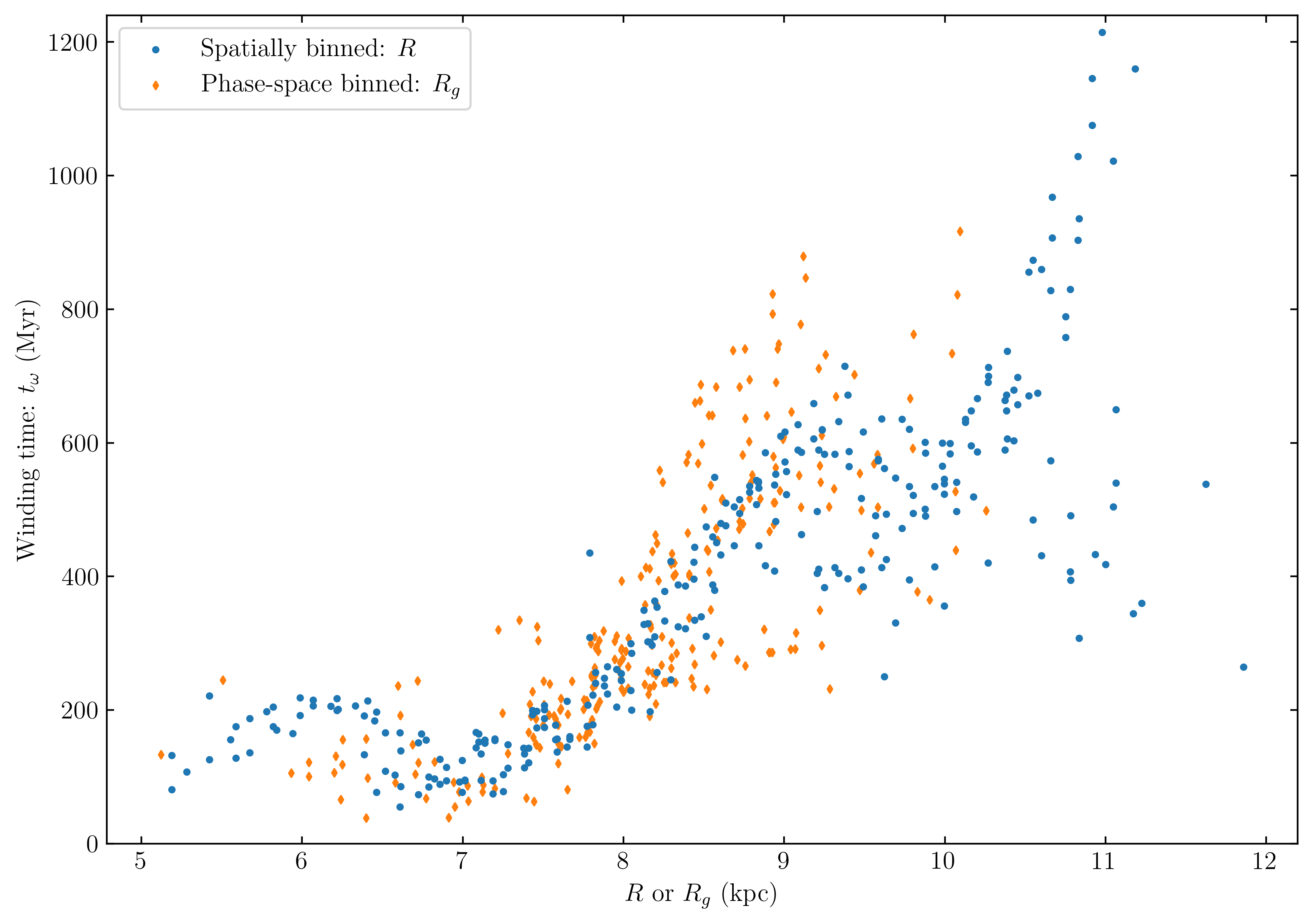}
    \caption{Winding time for the spatially binned data samples (as a function of radius $R$; circular markers) and the phase-space binned data samples (as a function of guiding radius $R_g$; diamond markers). We note that many data samples at small ($R\lesssim7~\kpc$) and large ($R\gtrsim11~\kpc$) radii were marked as dubious in our by-eye inspection. At high radii, there are seven outlier data samples with very high winding times, up to 1.9~Gyr, which are not included in this figure.}
    \label{fig:winding_time}
\end{figure*}

The structure of the simple model in Figure~\ref{fig:winding_rot_toy_model} do not match our observations. Instead, the data shows that both $\omega$ and $\varphi_{600}$ are close to constant with respect to $R$. Spatial variations in $\Gamma$ can affect $\omega$, but likely only by a few ten per cent.\footnote{As an example, in \texttt{gala}'s \texttt{MilkyWayPotential2022} \citep{gala}, $\Gamma$ varies with $R$ in an almost linear manner, taking values of 0.407 at $R=6~\kpc$ and 0.316 at $R=11~\kpc$. Since $\Gamma$ is higher valued at low $R$, accounting for this would make the slope seen in Figure~\ref{fig:winding_time} even steeper.}

As discussed above, around Eqs.~\eqref{eq:winding_morph2phys} and \eqref{eq:phase_angle_morph2phys}, self-gravity effects acting only in the disk's vertical dimension could delay $\omega$ but not simultaneously $\varphi_{600}$. For this reason, in the context of one-dimensional vertical dynamics, a uniform perturbation time cannot be reconciled with the flat radial profiles in $\omega$ and $\varphi_{600}$. Perhaps it can be explained in the context of a perturbing satellite, but this would require more complex, three-dimensional self-gravitating dynamics.

In Figure~\ref{fig:winding_time}, we show inferred winding times ($t_\omega$) for the data samples binned in space and in phase-space. We calculate $t_\omega$ assuming a disk surface density that decays according to the exponential function shown in Figure~\ref{fig:scatter_potscaling_XY}. For the spatially binned samples, we approximate them as being on circular orbits, giving
\begin{equation}
\label{eq:XY_winding_time}
    t_\omega = \omega \times \exp\Big( \frac{R-R_\odot}{2.6~\kpc} \Big) \times \Big(71.9~\frac{~\mathrm{Myr}}{\mathrm{rad}} \Big).
\end{equation}
The numerical constant is given by time per radian of winding, for the fiducial gravitational potential at $R=R_\odot$. For the phase-space binned data samples, we evaluate $t_\omega$ in the following way. First, we calculate the vertical energy for the two anchor heights (300 and 800~pc) using the vertical gravitational potential (where the mid-plane value is always zero) at the data sample's present-day radius ($R$), giving $E_{z,A}$ and $E_{z,B}$. Under the assumption that the vertical energy is conserved during a star's epicyclic motion, we then calculate the vertical period for $E_{z,A}$ and $E_{z,B}$ using the potential at the data sample's guiding radius ($R_g$), giving $P_{z,A}$ and $P_{z,B}$. The winding time is equal to $t_\omega = \omega / [2\pi(P^{-1}_{z,A}-P^{-1}_{z,B})]$. The phase-space binned data samples have a significantly higher scatter in the inferred $t_\omega$, due to the lower number of statistics and possibly factors pertaining to their varied non-circular orbits. If we instead assume that $J_z$ or $z_\mathrm{max}$ is conserved during a star's epicyclic motion, we get a very similar result, differing only by a few per cent on average.

\section{Discussion}
\label{sec:discussion}

In this work, we have studied the varying properties of the phase spiral, using either a spatial binning (cuts in $X$ and $Y$), or a four-dimensional phase-space binning (cuts in $\{X,Y,v_\phi,v_R\}$). Using the \emph{Gaia} DR3 proper motion sample, we have been able map the spiral properties to large spatial distances, out to $4~\kpc$ from the solar position. In the nearby spatial region, we have mapped the spiral properties in terms of the data samples' disk-plane parallel orbits, and found significant and complex structure. When comparing these two manners of binning, we see an overall strong agreement.

\subsection{Findings}
We summarize our most important findings in the list below.
\begin{itemize}
    \item The properties of the phase spiral vary most significantly with radius ($R$ or $R_g$), but also have complex and largely non-separable structure with respect to other disk plane parallel phase-space coordinates, as seen in Figures~\ref{fig:spat_plane_triple} and \ref{fig:XYUV_triple}. In our spatial binning, we see azimuthal dependence. In our phase-space binning, we see significant structure with respect to epicyclic motion, both in terms of epicyclic action and phase.
    \item The spiral properties of winding ($\omega$) and rotation phase ($\varphi_{600}$) vary smoothly across the studied volume (see Figures~\ref{fig:spat_plane_triple} and \ref{fig:XYUV_triple}), which covers a range of roughly 6~kpc in $R$ and nearly 8~kpc in $Y$. Furthermore, at the time that the perturbation was likely produced, many 100~Myr ago, the respective data samples were at opposite ends of the Milky Way, roughly 16 kpc apart (see e.g. Figure~\ref{fig:rewind_amplitude}).
    \item There are small-scale spatial variations in winding ($\omega$) that seem to correlate with the Milky Way's spiral structure, as seen in the middle panel of Figure~\ref{fig:spat_plane_triple}. In particular, there is a region of high winding which coincides with location of the Local Arm. We see a similar feature of high $\omega$ at $R_g \simeq 9~\kpc$ for the phase-space binned data samples.
\end{itemize}

\subsection{Interpretations}
These findings lead directly to some interpretations and conundrums. 
\begin{itemize}
    \item The uniformity of $\omega$ and in particular $\varphi_{600}$ indicate that the phase spiral was primarily sourced by one or many global perturbations. In a scenario of many spatially small-scale perturbations, local disk properties would set the balance of sourcing and dissipating phase spiral perturbations, which potentially could give rise to a smoothly varying profile for $\omega$. However, such stochastic perturbations are not expected to be correlated in terms of $\varphi_{600}$.
    \item The flat profile in $\omega$ is curious, since spiral winding is much faster in the heavier inner disk. As seen in Figure~\ref{fig:winding_time}, the inferred winding time varies significantly with radius, from roughly 150~Myr at $R=7~\kpc$ to 600~Myr at $R=9~\kpc$. The inferred winding times are sensitive to the disk scale length and possible radial variations in vertical anharmonicity ($\Gamma$), but such effects are comparatively small and can only marginally affect the radial slope.
    \item The uniformity of $\varphi_{600}$ is also consistent with a winding time that has a significant radial dependence. As a simplistic example, if we assume that the phase spiral was sourced by a global perturbation with a winding time that is inversely proportional to the disk surface density, then we would expect flat profiles in both $\omega$ and $\varphi_{600}$. As a counterexample, if we assume a uniform winding time, then we expect a very significant radial dependence for $\varphi_{600}$, as illustrated in the bottom panel of Figure~\ref{fig:winding_rot_toy_model}.
    \item It seems likely that our maps of spiral properties, in particular in terms of winding time, need to be explained in the context of our Galaxy's three-dimensional self-gravitating effects. A uniform perturbation time cannot be reconciled with our results if we consider self-gravity acting only in the disk's vertical dimension, whereby a $(z,w)$-dipole would spin without winding (i.e. with fixed rotating pattern, like a bar). Even if winding is significantly delayed in this scenario, the rotation phase would still evolve with a speed ($\de \varphi_{600} / \de t$) that is proportional to the disk surface density; this would give rise to a strong radial slope for $\varphi_{600}$, which is not observed.
    \item There are small-scale features in $\omega$ which seem to correlate with the Milky Way spiral arms, most clearly with the Local Arm at $R\simeq9~\kpc$. This could be explained by variations in the vertical gravitational potential, most importantly in terms of its anharmonicity ($\Gamma$), which could be amplified, for example, by a higher surface density of cold gas (see Appendix~\ref{app:grav_pot} for further details).
\end{itemize}

\subsection{Comparison with prior work}

\subsubsection{Data analyses}

The close-to-flat radial profiles that we observe for winding ($\omega$) and rotation phase ($\varphi_{600}$) are qualitatively consistent with the results of \cite{Antoja2023}, who used the \emph{Gaia} DR3 $\vlos$ sample, primarily binned in $R_g$, to analyze the phase spiral using an edge detection algorithm. In their figure 6, it is clear that both winding and rotation phase are close to constant with respect to $R_g$ in the studied range of roughly 5--11~kpc. In their figure 8, they see a significant slope of winding time with respect to $R_g$, as well as a smaller-scale feature of high winding close to $R_g\simeq 9~\kpc$. The slope we infer in this work is more dramatic, although a direct comparison is difficult given differences in data binning and modeling.

\cite{Frankel2023} used the \emph{Gaia} EDR3 $\vlos$ sample, binned in $R_g$ and $J_R$, and found significant and non-separable structure for the phase spiral's amplitude and winding time. Comparing with our phase-space binned results, the spiral amplitude has a similar overall slope with respect to $R_g$. They also see a smaller-scale feature of high winding at $R_g\simeq 9~\kpc$, but otherwise infer a close-to-constant winding time. We have not been able to find a convincing reason for this winding time discrepancy. It could be due to, at least in part, differences in data binning and modeling. Our results demonstrate that the epicyclic phase is important, which \cite{Frankel2023} does not account for, although this seems unlikely to fully explain the discrepancy. Another factor could be related to the spiral's evolution as it traverses different $R$ during its epicyclic motion; however, when we assume a conserved $J_z$ (as do \citealt{Frankel2023}), we obtain a practically identical results as when we assume a conserved $E_z$. Since we see consistent results for our spatial binning and phase-space binning schemes, we consider our general result to be robust.

\cite{Darragh-Ford2023} used their spiral-fitting \texttt{ESCARGOT} algorithm with \emph{Gaia} DR3, binned in guiding radius ($R_g$) and azimuth ($\phi$). Their inferred winding times, shown in their figure 8, are generally consistent with our results, with an overall slope that rises from roughly 350~Myr at $R_g \simeq 7.5~\kpc$, to 600~Myr at $R_g \simeq 10~\kpc$, and with a smaller-scale feature of high values at $R_g \simeq 9.3~\kpc$.

\cite{Alinder2023} studied the phase spiral in \emph{Gaia} DR3, with a focus on its amplitude and rotation phase in the outer disk. They found a positive correlation between the rotation phase with respect to Galactic azimuth, at a rate of roughly $3^\circ$ per $1^\circ$. This is consistent with our results for the spatially binned data samples, as shown in Figure~\ref{fig:spat_plane_triple}.
\cite{Darragh-Ford2023} show results for the spiral rotation phase in \emph{Gaia} DR3, in their figure 8. However, a direct comparison is difficult since they use a different parametrization; they present to rotation angle for a fixed vertical frequency without explicitly stating the assumed gravitational potential used to calculate those frequencies.

We infer a disk scale length of 2.9~kpc, as seen in Figure~\ref{fig:scatter_potscaling_XY}, which is in agreement with the general consensus. A review by \cite{BlandHawthorn2016} summarizes 15 articles and reports $2.6\pm 0.5~\kpc$. Later analyses include $2.2\pm 0.1~\kpc$ \citep{Widmark-spiral-III}, $2.4 \pm 0.1~\kpc$ \citep{Wang2022}, roughly 3.9~kpc \citep{Robin2022}, $2.17^{+0.18}_{-0.08}~\kpc$ \citep{Ibata2023}, and 3.3--4.2~kpc \citep{WidmarkNaik2024}. Variations between results could be due to deviations from the assumed axisymmetric and exponentially decaying disk profile, which can bias the results in different ways depending on the chosen tracer sample and spatial volume.

\subsubsection{Simulations}

We can compare our results for winding (or winding time) as a function of radius, with predictions coming from simulations. \cite{Darragh-Ford2023} tested their spiral-fitting \texttt{ESCARGOT} method on a test-particle simulation. They inferred the spiral's winding time in different disk locations, in order to compare with the passage of a perturbing satellite at $t\simeq-840~\Myr$. They found similar winding time values across the disk, typically varying only a few 10~Myrs for well-fitted spirals.

For self-consistent simulations, there are few literature sources that show explicit plots of inferred winding or winding time as a function of radius. \cite{Darragh-Ford2023} also applied \texttt{ESCARGOT} to a high-resolution simulation from \cite{Hunt2021}, in a small spatial volume analogous to the solar neighbourhood, for three guiding radius bins. As shown in their figure 10, the inferred winding times do not line up with the satellite's pericenter passages, probably as an effect of self-gravity. There is no strong dependence on guiding radius, although the outermost bin has a shorter winding time (contrary to the results of this work). \cite{Widmark-spiral-IV} also inferred winding times for the same simulation, at different spatial radii and azimuths. Their results, found in their appendix, are somewhat noisy and poorly resolved; there is no strong general trend, but for some fixed galactic azimuths there are indeed significant slopes, both positive and negative, for the winding time as a function of radius. However, it is difficult to make a fair comparison given that the simulation from \cite{Hunt2021} is not a perfect analog of the Milky Way and Sagittarius satellite (e.g. the simulated satellite is likely too massive). \cite{Asano2025} studied the phase spiral in simulations with self-gravity. In their figure 10, they show the phase spiral in bins of $R$ and $\phi$, roughly 570~Myr after a massive satellite passage. Although they did not estimate or model the winding time, it is clear that the inner disk (low $R$) has a higher degree of winding than the outer disk (high $R$).

\cite{Tepper-Garcia2025} have run self-consistent simulations with an interstellar medium component. They find that the presence of turbulent gas, driven by stellar feedback, gives rise to phase spiral properties that have a high degree of intermittency on kilo-parsec scales. It is difficult to make a direct comparison with our results; for example, \cite{Tepper-Garcia2025} study the spiral in $v_\phi$ and $v_R$, as opposed to stellar number density, and use significantly larger spatial bins. While we do observe intermittency, in particular for phase spiral amplitude and winding, these structures seem comparatively weak, setting an upper limit to the influence that gas turbulence has over the phase spiral's evolution in the Milky Way.

\section{Conclusion}
\label{sec:conclusion}

We have studied the phase spiral in the Milky Way disk, using data from \emph{Gaia} DR3, supplemented with spectro-astrometric parallax estimates and line-of-sight velocity predictions from Bayesian Neural Networks. We have produced high resolution maps of the spiral's present-day morphological properties, as a function of disk plane parallel phase-space coordinates.

We see uniformity in the spiral's winding and rotation phase over the studied disk area, which is several kilo-parsecs wide in both Galactic azimuth and Galactocentric radius. Especially the uniformity in rotation phase is evidence that the spiral was primarily sourced by one or many global perturbations, as opposed to local perturbations driven by small scale Galactic structure, for example from dark matter sub-halos.

These results are highly informative of the phase spirals origin and evolution, and we can already draw strong conclusions. However, they also present conundrums, in particular in terms of the radial trends for the inferred winding time. Going forward, the data features we have identified are important benchmarks that spiral models and simulations should be tested against.


\section*{Acknowledgments}

This work made use of the following open-source Python packages: \textsc{Matplotlib} \citep{matplotlib}, \textsc{NumPy} \citep{numpy}, \textsc{SciPy} \citep{scipy}, \textsc{Pandas} \citep{pandas}, \textsc{TensorFlow} \citep{tensorflow2015-whitepaper}. AW is supported by the European Union's Horizon 2020 research and innovation program, under the Marie Skłodowska-Curie grant agreement number 101106028. JH acknowledges the support of a UKRI Ernest Rutherford Fellowship ST/Z510245/1. KVJ and KT's contributions were enabled by a Simons Foundation award 1018465.

\bibliography{refs}{}
\bibliographystyle{aasjournal}



\appendix

\section{Solar neighborhood model of the vertical gravitational potential}
\label{app:grav_pot}

\begin{figure*}
    \centering
    \includegraphics[width=0.56\textwidth]{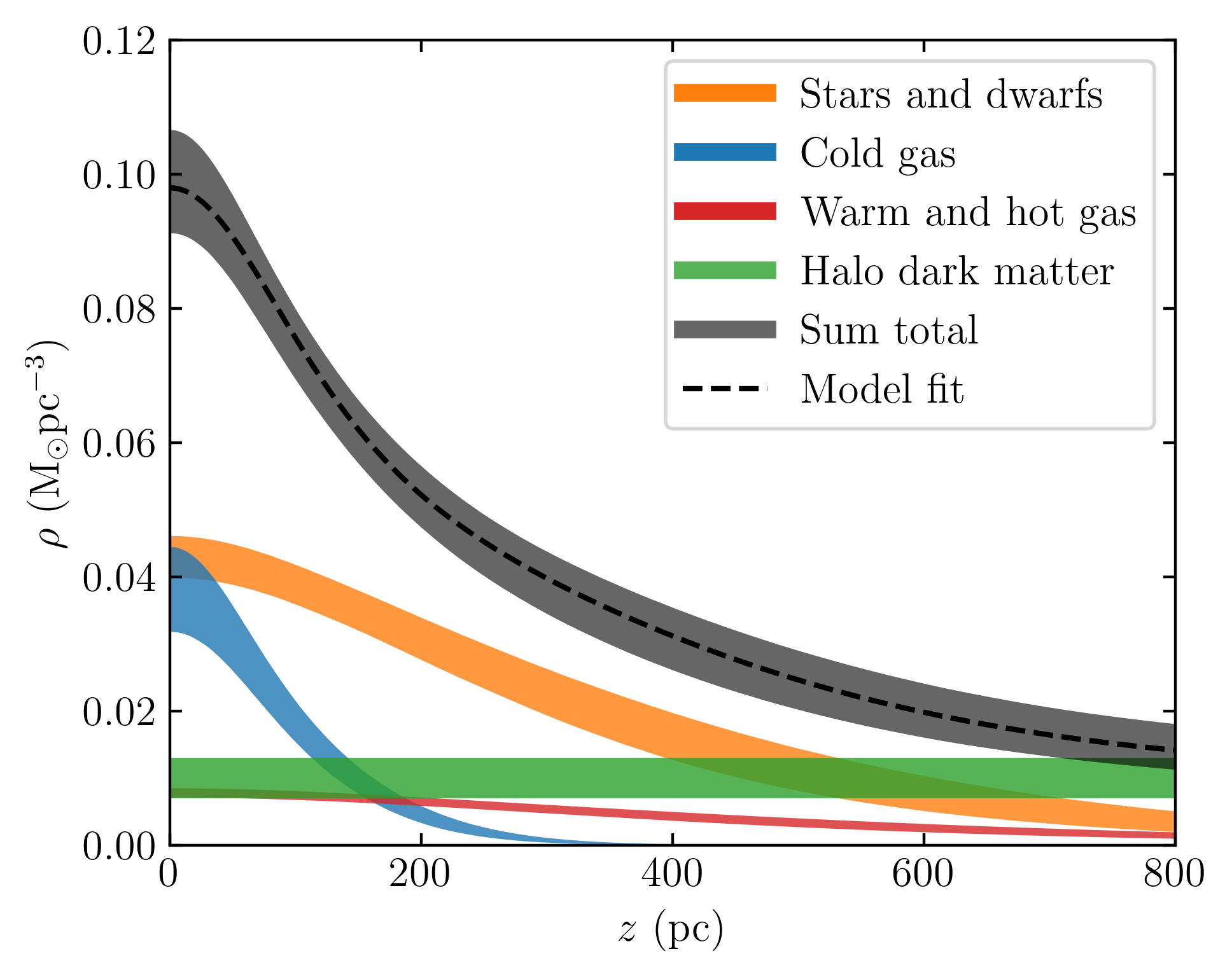}
    \caption{The matter densities in our solar neighborhood model for $\Phi_\odot(z)$, based on results cataloged by \cite{Schutz2017}. The mass components are split into stars and dwarfs, cold gas, warm and hot gas, and dark matter. The sum total is shown in gray. The dashed black line shows our fitted functional form.}
    \label{fig:rho_model}
\end{figure*}

As described in Section~\ref{sec:vertical_orbits}, we model the vertical gravitational potential as being proportional to a solar neighborhood model, written $\Phi_\odot(z)$. Under the assumption of vertical separability, the vertical gravitational potential is related to the total matter density distribution, written $\rho_\odot(z)$, through the one-dimensional Poisson equation. We use a solar neighborhood model based on direct observations of baryonic matter density components (as opposed to results from a dynamical mass measurement). This model was compiled by \cite{Schutz2017}, using results from \cite{Flynn2006,McKee2015,Kramer2016}. We fit a functional form to this model, assuming a local dark matter density of $0.01\pm0.003~\Msunppcc$ \citep{Salas2021}, given by
\begin{equation}
\begin{split}
    & \rho_\odot(z) =
    (0.01~\Msunppcc) \\
    & + (0.043~\Msunppcc) \, \text{cosh}^{-2}\bigg(\frac{z}{120~\pc}\bigg) \\ 
    & + (0.045~\Msunppcc) \, \text{cosh}^{-2}\bigg(\frac{z}{430~\pc}\bigg).
\end{split}
\end{equation}

The solar neighborhood model and functional fit are shown in Figure~\ref{fig:rho_model}. The density close to the disk mid-plane is roughly equal parts stars and gas. Cold gas is the most uncertain component, which is further subdivided into atomic and molecular gas (HI and H$_2$). Molecular gas is particularly difficult to observe, since it lacks a permanent electric dipole moment. Typically, carbon monoxide (CO) observations are used as a tracer of H$_2$ \citep{Heyer2015}, although the precise conversion factor, as well the amount of CO-dark H$_2$, are poorly constrained \citep{Grenier2005,Wolfire2010,Bolatto2013,Tang2016,Reach2017,WidmarkGamma2023}. For these reasons, the cold gas component could suffer systematic biases that are larger than the model's reported statistical uncertainty. Furthermore, the cold gas is highly structured \citep{Kalberla2009,Heyer2015}, and could vary significantly between the different disk regions studied in this work.

The amount of cold gas is important for the mid-plane density and level of anharmonicity in the vertical gravitational potential, which affects how quickly the phase spiral winds up. For the solar neighborhood model used in this work, $\Gamma=0.26$ between our anchor heights of $z_1=300~\pc$ and $z_2=800~\pc$. If we remove the cold gas component completely, we obtain $\Gamma=0.17$. If we double the cold gas density, we obtain $\Gamma=0.32$. These are dramatic, but not entirely unrealistic, changes to the vertical gravitational potential model, giving an idea about the extent to which $\Gamma$ could vary between different disk regions.

\section{Test-particle simulation}
\label{app:sim_results}

\begin{figure*}
    \centering
    \includegraphics[width=0.5\textwidth]{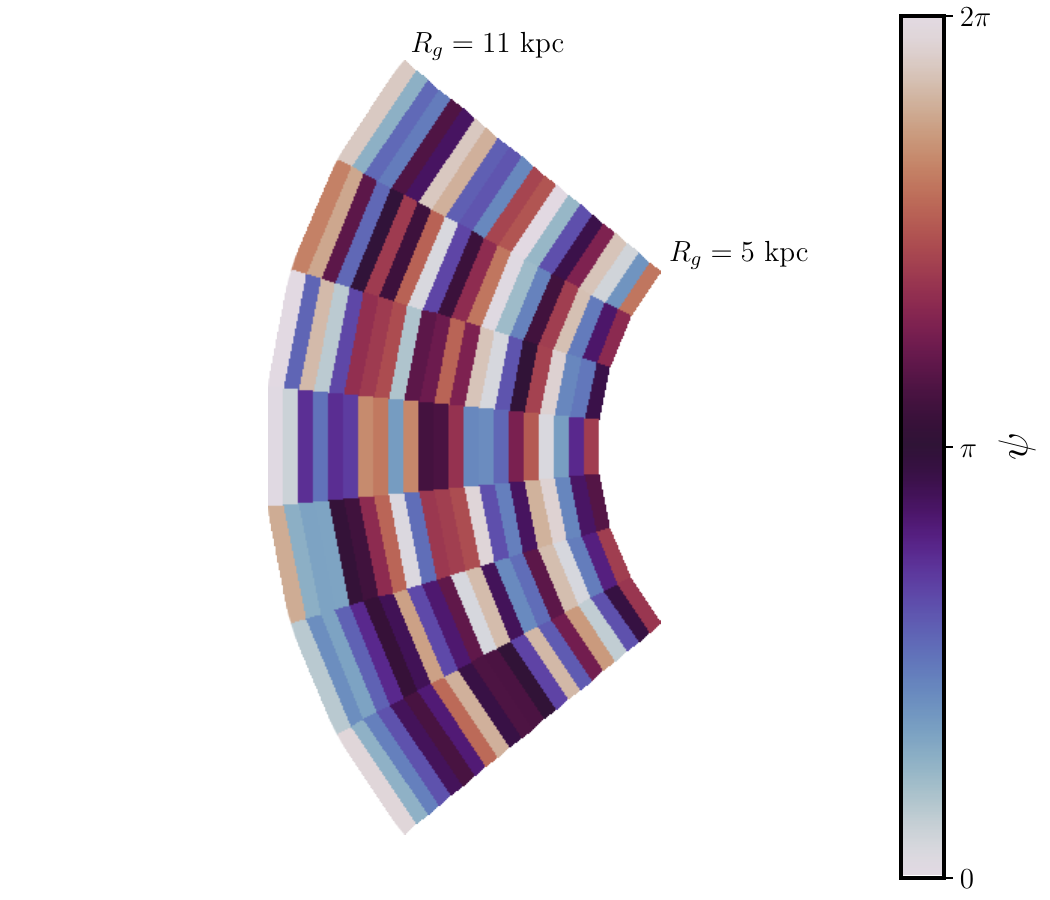}
    \caption{The spiral rotation phase for different disk regions in a test particle simulation (see Appendix~\ref{app:sim_results} for details). This shows the phase angle calculated 0.25 Gyr after the satellite passage at a vertical position of two times the scale height (at $R_g\sim 8.2 kpc$) of the simulated disk. The steep slope with respect to guiding radius, creating a sawtooth pattern through the mod $2\pi$ operator, is qualitatively the same as the simplistic model shown in Figure~\ref{fig:winding_rot_toy_model}. This is in contrast to our inferred results, seen for example in the right panel of Figure~\ref{fig:spat_plane_triple}.}
    \label{fig:test_part_sim_phase}
\end{figure*}

We compare our spiral rotation phase ($\varphi_{600}$) measurements in the data to those in a test particle simulation.
Our simulation contains a Milky Way-like host and a dwarf galaxy qualitatively similar to the Sagittarius dwarf galaxy.
The structure of both galaxies are represented by static, analytic functions. The simulation was run for 3~Gyr, with a time step of 10~Myr. As opposed to the real Sagittarius, this dwarf does a single flyby, crossing the disk with a Galactocentric radius of $\sim 16~\kpc$.

We focus on a similar region of the disk as in the observations (an angular sector corresponding to one quarter of the disk with $5~\kpc < R_g < 11~\kpc$).
We split this area into 154 regions (22 radial bins and 7 angular ones, shown in Figure~\ref{fig:test_part_sim_phase}) at each timestep and then use a Fourier-Laguerre basis to quantitatively describe the (action-angle) phase spirals in each region.
From this basis function expansion, we get a pitch angle for the spirals and can use it to derive the phase angle at any vertical height.
Because the test particle simulation is in action-angle coordinates, we cannot directly report the $\psi_{600}$ value defined exactly the same as for the the observed data.
Instead we provide the analogous result by calculating the rotation phase of each spiral at two times the scale height (at $\sim 8.2$ kpc) of the disk.

We show the result of this calculation in Figure~\ref{fig:test_part_sim_phase}.
The pattern is clearly in qualitative agreement with the the simplistic model shown in Figure~\ref{fig:winding_rot_toy_model}. They both have a steep slope with respect to radius, creating a sawtooth pattern through the mod $2\pi$ operator. As is evident from Figures~\ref{fig:spat_plane_triple} and \ref{fig:winding_rot_toy_model}, this slope is not seen in the observed data.

\section{Supplementary result plots}
\label{app:supp_results}

\begin{figure*}
    \centering
    \includegraphics[width=0.7\textwidth]{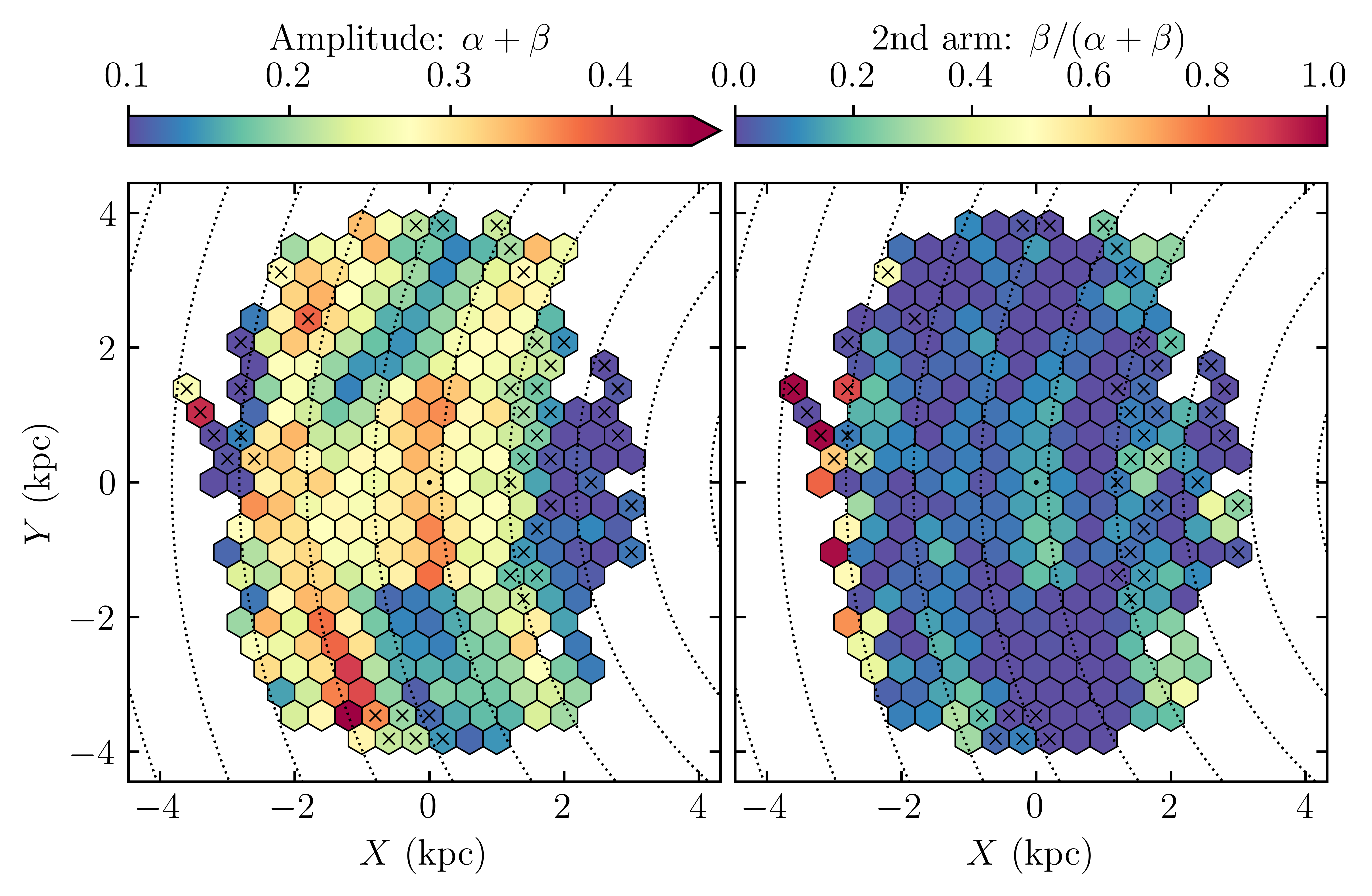}
    \caption{Analogous to Figure~\ref{fig:spat_plane_triple}, but here showing the phase spiral's relative over-density amplitude of the spatially binned data samples. The left panel shows the sum of the anti-symmetric and symmetric components ($\alpha+\beta$). The right panel shows the relative strength of the double-armed symmetric component ($\beta/(\alpha+\beta)$). We stress that these spiral parameters are very sensitive to spatially varying systematics; for example, we see a discontinuity that coincides with the line-of-sight velocity cut (see the black outline in Figure~\ref{fig:num_count}).}
    \label{fig:XY_double}
\end{figure*}

In this appendix section we show some supplementary results. In Figure~\ref{fig:XY_double}, we show the spiral amplitude parameters, which were not included in Figure~\ref{fig:spat_plane_triple} for the spatially binned data samples. We do see significant structure, in particular in terms of the total amplitude ($\alpha+\beta$), but this is likely mainly driven by strong selection effects.

\begin{figure*}
    \centering
    \includegraphics[width=0.7\textwidth]{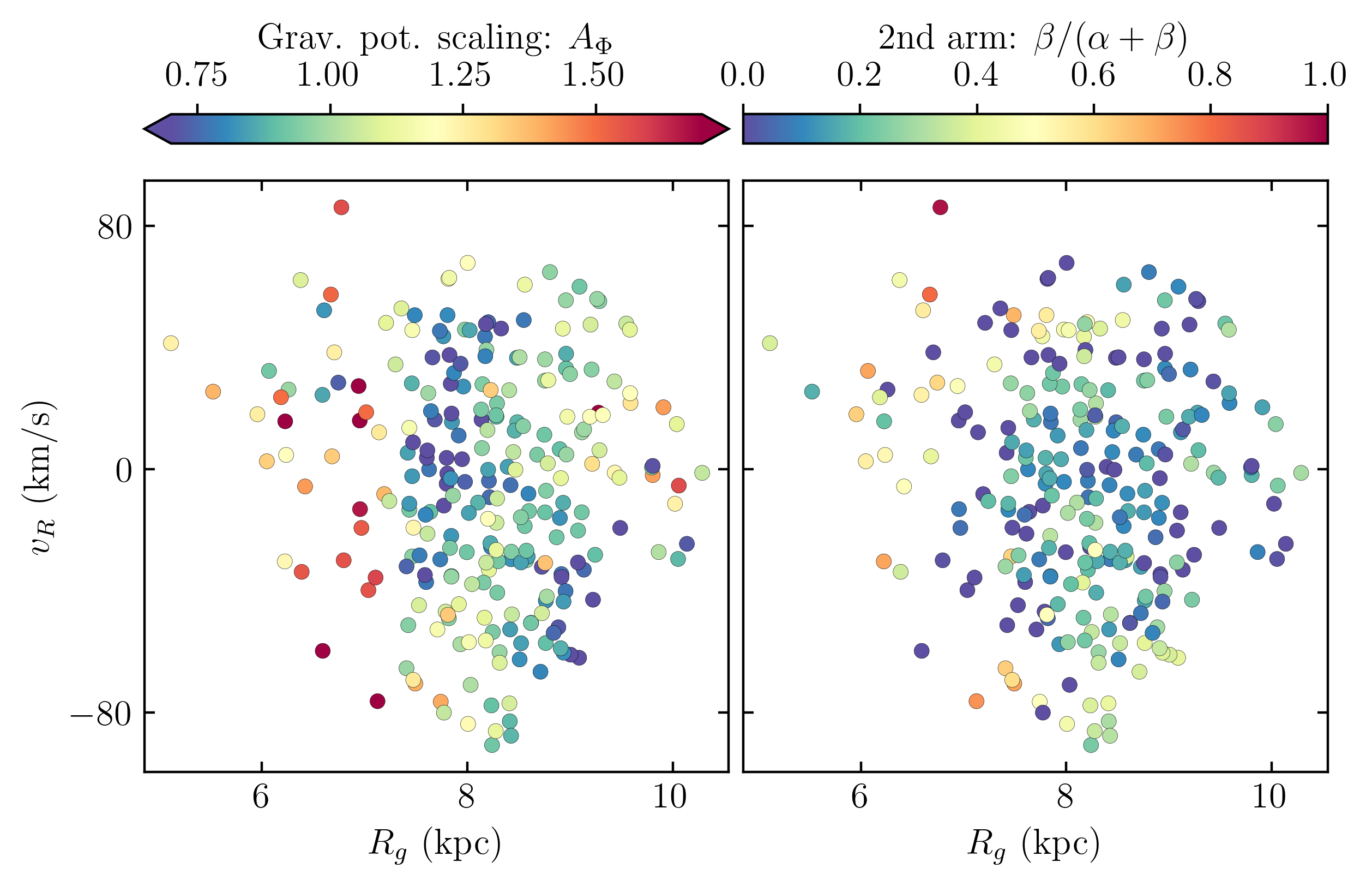}
    \caption{Analogous to Figure~\ref{fig:XYUV_triple}, showing inferred phase-spiral properties of the phase-space binned data samples. The left panel shows the gravitational potential scaling ($A_\Phi$). The right panel shows the relative strength of the double-armed symmetric component ($\beta/(\alpha+\beta)$).}
    \label{fig:XYUV_double}
\end{figure*}

\begin{figure*}
    \centering
    \includegraphics[width=0.5\textwidth]{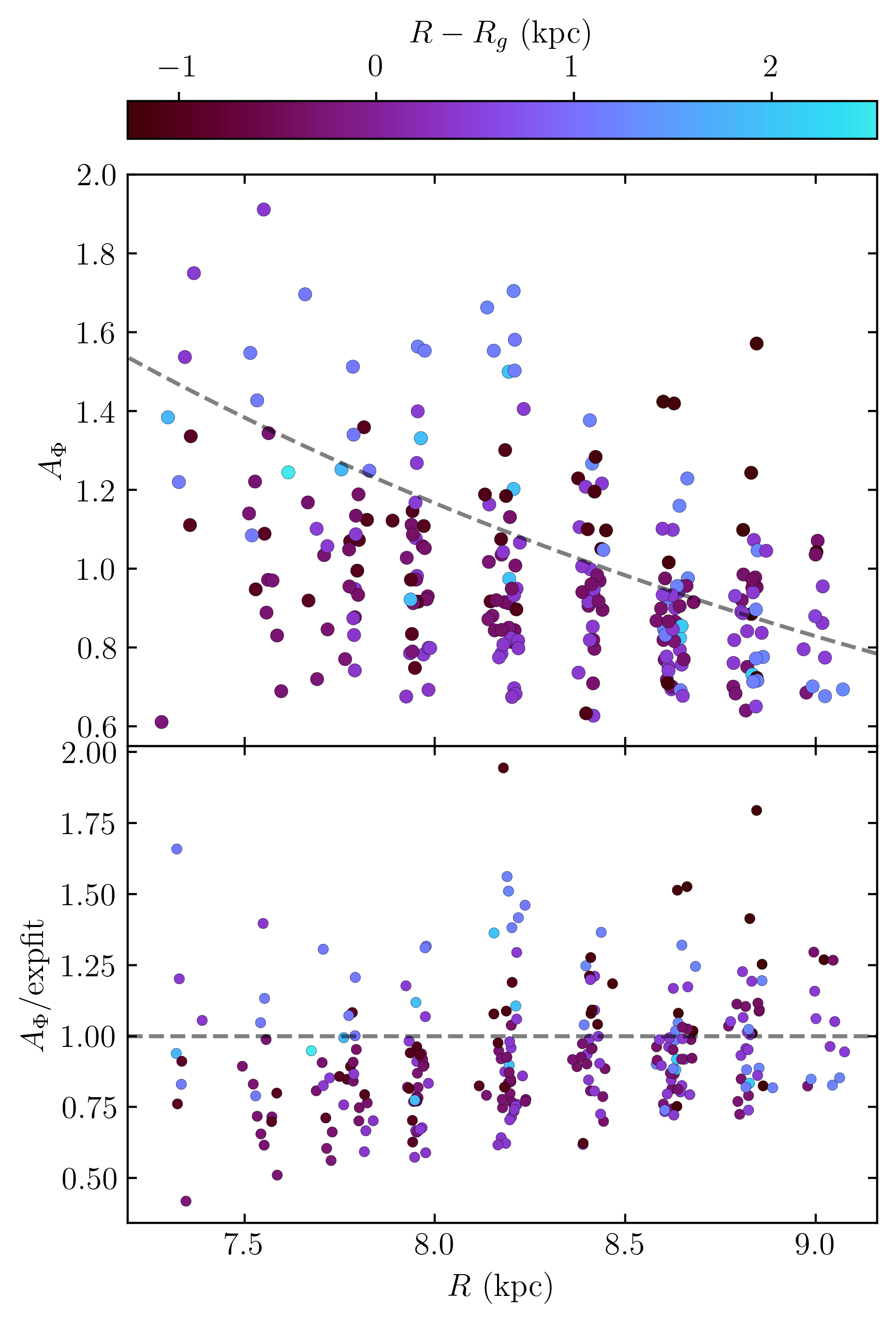}
    \caption{Analogous to Figure~\ref{fig:scatter_potscaling_XY}, but for the phase-space binned data samples. The scatter points are color coded by $R-R_g$, meaning the data sample's present-day epicyclic displacement. The dashed line is the same as shown in Figure~\ref{fig:scatter_potscaling_XY} (i.e. fitted to the spatially binned data samples), although the axis ranges differ.}
    \label{fig:scatter_potscaling_XYUV}
\end{figure*}

In Figure~\ref{fig:XYUV_double}, we show the remaining two spiral parameters ($A_\Phi$ and $\beta/(\alpha+\beta)$) of the phase-space binned data samples that were not already included in Figure~\ref{fig:XYUV_triple}. In Figure~\ref{fig:scatter_potscaling_XYUV}, we show a scatter plot of $A_\Phi$ as a function of $R$ for the phase-space binned data samples, analogous to Figure~\ref{fig:scatter_potscaling_XY}. The inferred values of $A_\Phi$ are generally consistent with the results of the spatially binned data samples as a function of $R$. However, there is clearly some structure that also depends on $R_g$ and $v_R$, indicating some degree of systematic bias. As seen in the right panel of Figure~\ref{fig:XYUV_double}, the relative strength of the symmetric spiral component (i.e., second arm) is generally small, with some higher values at low guiding radii (consistent with the findings of \citealt{Hunt2022}).

\begin{figure*}
    \centering
    \includegraphics[width=1.0\textwidth]{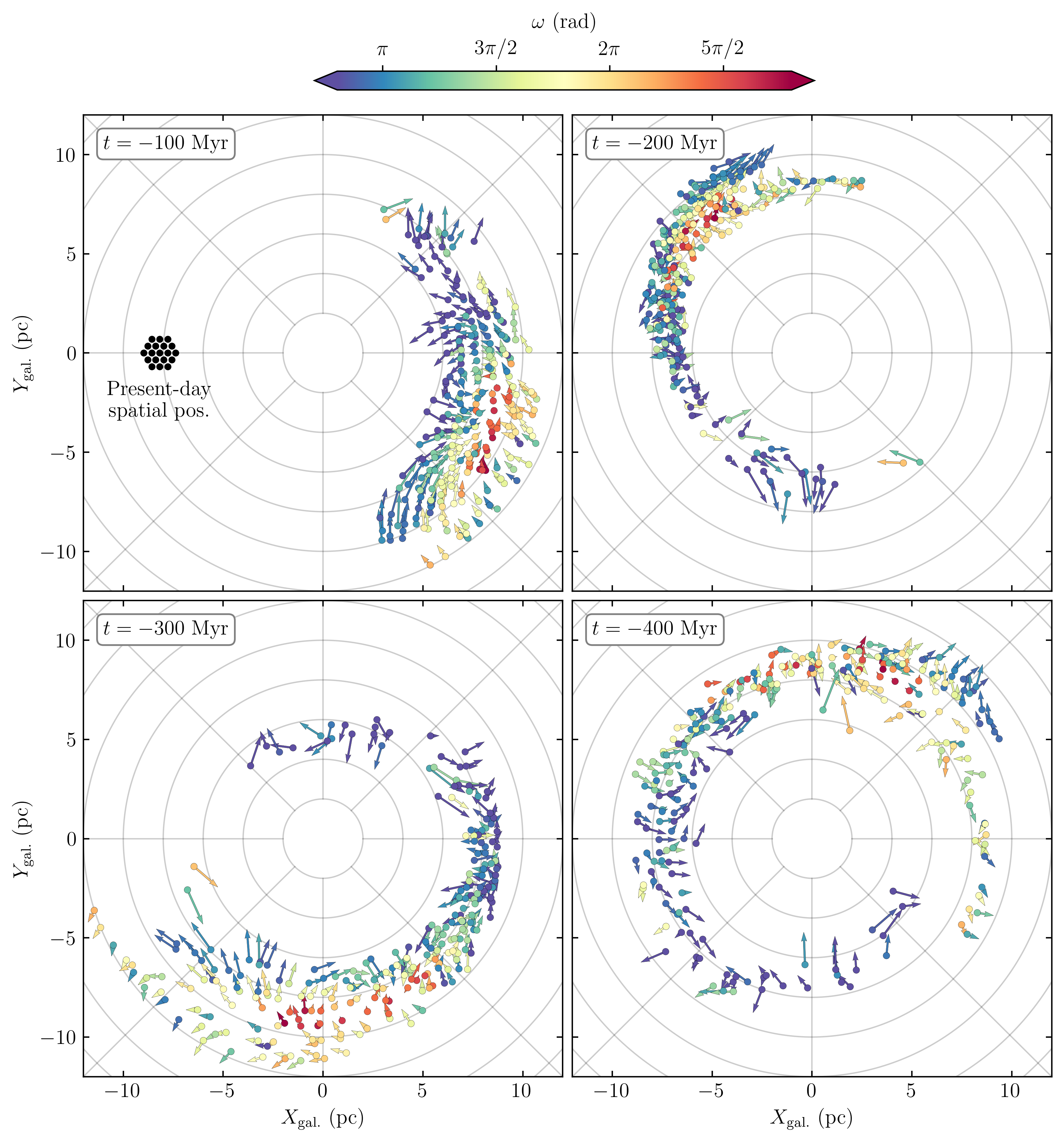}
    \caption{Same as Figure~\ref{fig:rewind_amplitude}, but for the winding parameter ($\omega$) of the phase spiral.}
    \label{fig:rewind_winding}
\end{figure*}

\begin{figure*}
    \centering
    \includegraphics[width=1.0\textwidth]{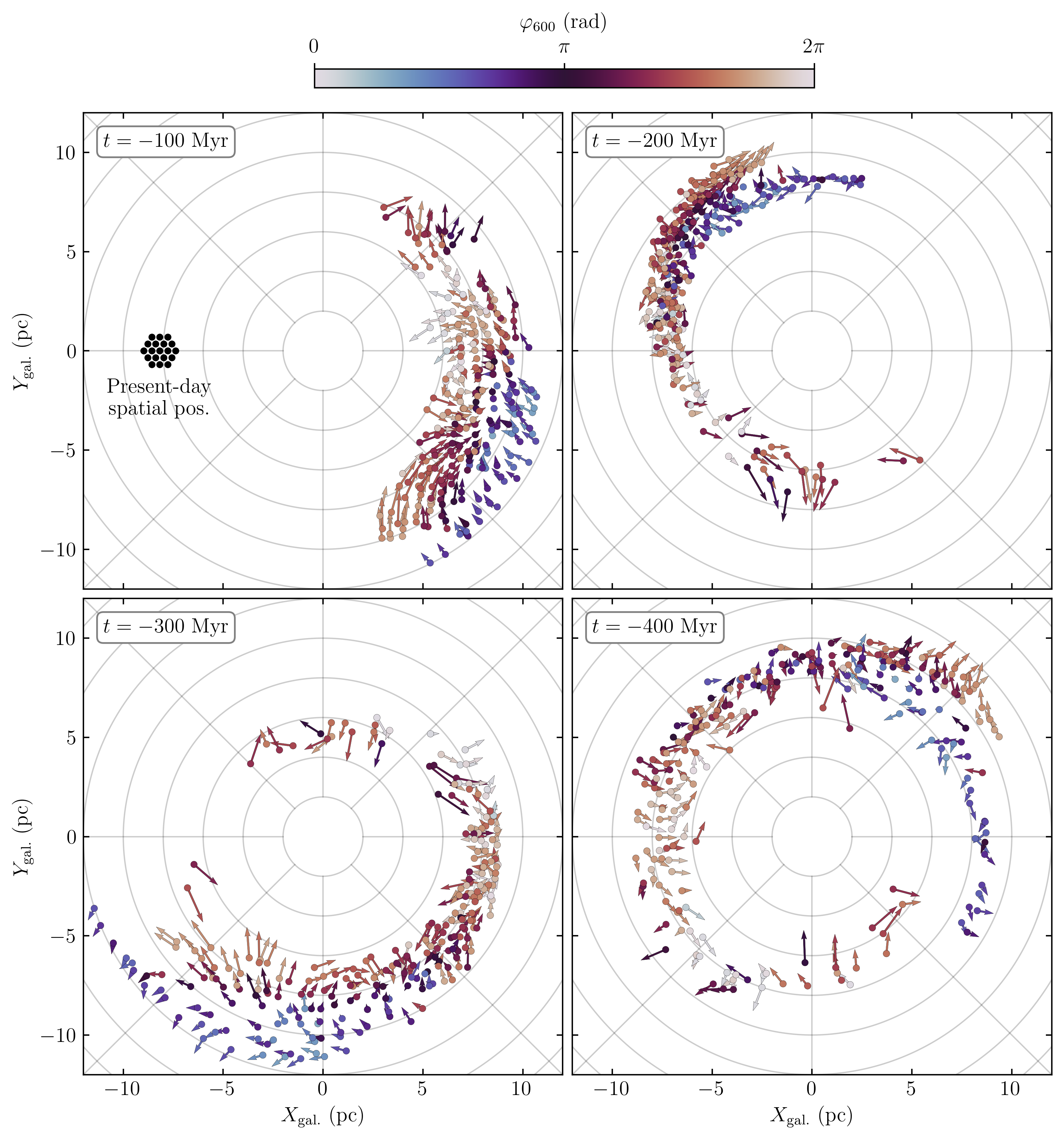}
    \caption{Same as Figure~\ref{fig:rewind_amplitude}, but for the rotation phase ($\varphi_{600}$) of the phase spiral. The colorbar is cyclical.}
    \label{fig:rewind_rotphase}
\end{figure*}

In Figures~\ref{fig:rewind_winding} and \ref{fig:rewind_rotphase}, we show the phase-space binned data samples in terms of their backward evolved spatial positions in the Galactic plane, analogous to Figure~\ref{fig:rewind_amplitude}, but for phase spiral parameters $\omega$ and $\varphi_{600}$. Is is clear that the region of high $\omega$ correspond to orbits that are close to circular. For some time snapshots, in particular $t=-300~\Myr$, the distribution of inferred $\varphi_{600}$ varies smoothly with spatial position.

\begin{figure*}
    \centering
    \includegraphics[width=.5\textwidth]{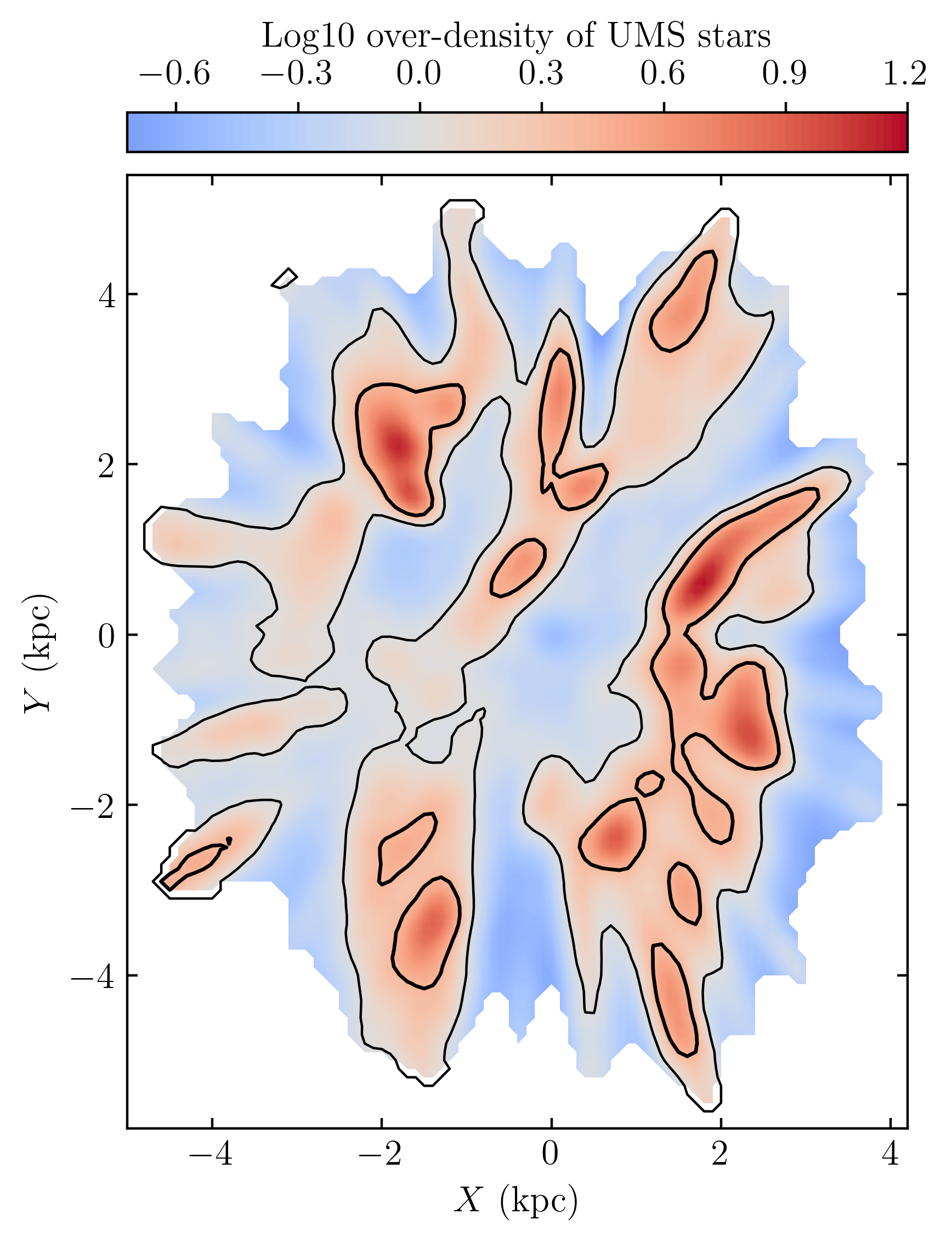}
    \caption{Over-density of upper main sequence stars in the disk plane, from \citet{Poggio2021} using \emph{Gaia} EDR3. The overlaid black lines are the same contour lines as in the middle panel of Figure~\ref{fig:spat_plane_triple}, corresponding to log$_{10}$ over-density values of 0 (thin line) and 0.4 (thick line).}
    \label{fig:poggio}
\end{figure*}

In Figure~\ref{fig:poggio}, we show the over-density of upper main sequence stars in the disk plane \citep{Poggio2021}, which serves as a tracer of Galactic spiral structure. This map is most likely less robust in the most distant spatial regions, in particular in the direction of the Galactic center. The observed structure seems to correlate well with the regions of high $\omega$ seen in the middle panel of Figure~\ref{fig:spat_plane_triple}. The Local Arm, corresponding to the over-dense band roughly 1~kpc outside the solar position, has been observed with other tracers and its nature is debated. For example, it is unclear if and how far it extends into negative azimuth angles (i.e., into negative $Y$, \citealt{Reid2014,Xu2016}).

\end{document}